\documentclass[%
 reprint,
 superscriptaddress,
 amsmath,amssymb,
 aps,
 prl,
]{revtex4-2}

\usepackage{physics}% for brakets
\usepackage{graphicx}% Include figure files
\usepackage{dcolumn}% Align table columns on decimal point
\usepackage{bm}% bold math
\usepackage{booktabs} % good looking tables
\usepackage{amssymb}

\usepackage{xcolor}
\usepackage{tikz}
\usetikzlibrary{calc,positioning,backgrounds,decorations.pathmorphing,arrows.meta,shapes.geometric,fadings,fit}

\definecolor{deeppurple}{HTML}{4D38F5} 
\usepackage{hyperref}
\hypersetup{colorlinks=true,allcolors=deeppurple}

\begin{document}
\title{Reference-free certification of machine-learning interatomic potentials}

\author{Jonas H{\"a}nseroth}
\email{jonas.haenseroth@tu-ilmenau.de}
\affiliation{Theoretical Solid State Physics, Institute of Physics, Technische Universit{\"a}t Ilmenau, 98693 Ilmenau, Germany}

\author{Christian Dre{\ss}ler}
\affiliation{Theoretical Solid State Physics, Institute of Physics, Technische Universit{\"a}t Ilmenau, 98693 Ilmenau, Germany}

\date{\today}

\begin{abstract}
Universal machine-learning interatomic potentials now reach held-out energy and force errors so small that they no longer predict how a model behaves in simulation.
Here we show that a potential can be graded without any reference calculation, against properties the exact Born-Oppenheimer surface satisfies by mathematical or physical necessity.
We organise these properties into four families, symmetry and invariance, self-consistency and integrability, statistical-mechanical equilibrium and regularity, and turn them into fourteen inexpensive probes, each certifying against a target that is exact and independent of chemistry and reference method.
Every probe therefore returns an absolute, architecture-comparable score from the energies, forces and stresses a potential already exposes, and reveals failures a fixed test set cannot.
Applied to $64$ pretrained potentials, the suite resolves two orders of magnitude of certified quality at comparable reported accuracy, and flags a fine-tuned potential whose reference errors improve while its surface degrades until molecular dynamics fails.
An interactive leaderboard of all 64 models is available at \url{https://jhaens.github.io/pescert-bench}.
\end{abstract}

\maketitle

Machine-learning interatomic potentials (MLIPs) have advanced rapidly: models trained on large and chemically diverse datasets can now be applied across much of the periodic table~\cite{macemp0,mattersim,orb,uma,dpa,pet_oam}, reaching near-ab initio accuracy in energies and forces on held-out structures~\cite{forces-are-not-enough,mliparena,matbench}.
These numbers are necessary but increasingly insufficient: two models with indistinguishable test-set errors can behave very differently in a molecular-dynamics run and a model can attain excellent errors while failing to recover material-specific behaviour such as reactive or rare events~\cite{atk,atk_found,atk_size}.
A useful litmus test makes the gap concrete: can a model obtain a perfect energy and force error on a test set and still fail the check?
Whenever the answer is yes, the check measures something the training loss does not.
The failures that matter in practice, a trajectory that slowly heats under a thermostat, an optimisation that drifts off a symmetric configuration, a variable-cell run whose angles slowly shear away, are precisely of this kind: they are invisible to an average error over a fixed set of snapshots, yet they decide whether a simulation is trustworthy.

A second difficulty is that most physically motivated checks require a first-principles reference at the desired level of theory: phonon spectra, optimised ground or transition states, or energy barriers compared against density-functional theory.
Community benchmarks and leaderboards likewise evaluate MLIPs almost exclusively by how well they recover such reference properties~\cite{matbench,mliparena,lambenchbenchmark,mlipstudio,mlipaudit,jarvisleaderboard}.
Such comparisons inherit the functional and basis-set ambiguity of the reference, are vulnerable to data leakage when that reference overlaps the training set and cannot be placed on a common absolute scale across heterogeneous models, since a chemistry-specific target such as the rate constant of an organic reaction is meaningless for a model trained solely on inorganic crystals.
They also conflate two different questions, whether the model reproduces a particular level of theory and whether the model is internally consistent as a potential at all, so that a poor score cannot be attributed unambiguously to either cause.

Here we pursue the opposite strategy.
We identify properties that the true potential-energy surface satisfies exactly, by mathematical or physical necessity, for every system regardless of chemistry and that can be certified against an exact number without ever computing a reference.
The exactness of the target is the crucial ingredient: a probe whose ideal value is exact has no functional or basis-set ambiguity, no data leakage and a meaningful absolute scale, which is precisely what allows a single score to be compared across wildly different models.
We show that a small number of such identities, drawn from symmetry, integrability, statistical mechanics and regularity, already covers the behaviours that matter most in practice and that they are cheap enough to run in minutes per model on commodity hardware.

Only a small number of previous studies has discussed related evaluations, such as the smoothness of the predicted potential-energy surface~\cite{bsct} probed by elongating and shortening bonds in arbitrary systems, the equivariance of the predicted outputs~\cite{pet_oam} and the influence of conservative versus non-conservative force prediction on overall model quality as well as on molecular-dynamics stability and energy conservation~\cite{darkside}.

\begin{figure*}[ht]
    \centering
    \includegraphics[width=0.9\textwidth]{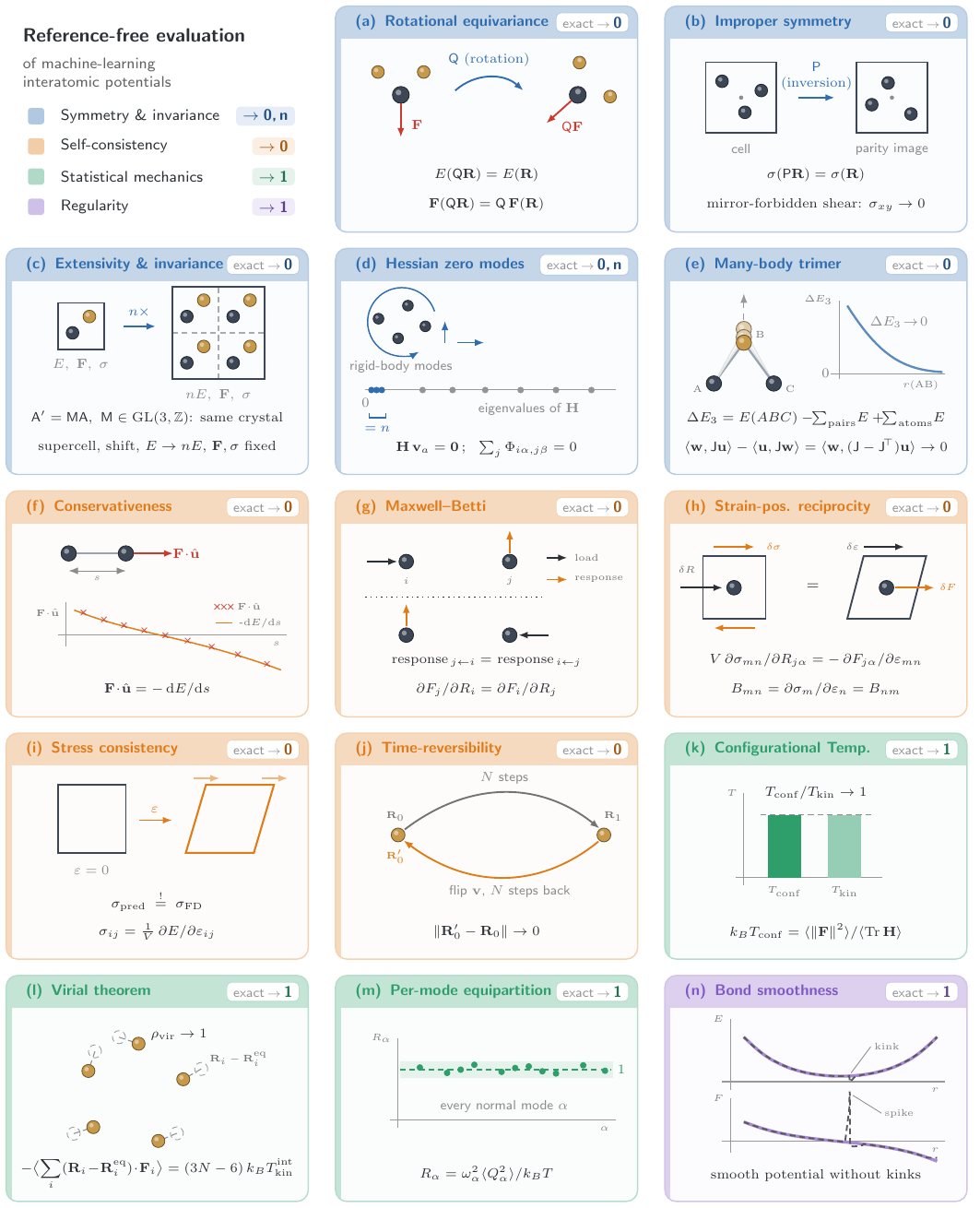}
    \caption{
    \textbf{Reference-free certification probes.}
    Each panel states one identity that the exact PES should satisfy and the exact value it certifies against.
    Panel (n) reproduces the potential-smoothness evaluation introduced by Liu et al.~\cite{bsct}.
    }
    \label{fig:overview}
\end{figure*}

The contribution of this work is threefold.
First, we give a unifying account of reference-free identities and organise them into four families (Figure~\ref{fig:overview}).
Second, we turn these identities into a concrete, uniform suite of probes, several of them, to our knowledge, not previously used to grade MLIPs, each returning the same result object: an exact target, a non-negative deviation from it, a normalised score and a model-call budget.
Third, we provide an open, architecture-independent implementation that requires only the energies, forces and (optionally) stresses any potential already exposes and use it to compare $64$ contemporary large-scale pretrained models.

\section*{Certifications}

Every probe we use represents a property the true Born--Oppenheimer surface satisfies by necessity, that holds for every system and that is certifiable against an exact number; then measure how far the model deviates from that number.
We summarise the four families here and detail the individual probes, together with exactly how each deviation is computed below.

The first family is symmetry and invariance.
The energy is invariant under global translations, rotations, reflections and inversion and permutations of like atoms and the forces and stress transform equivariantly~\cite{mace,pet_oam}.
The same invariance extends to the bookkeeping of the periodic description: the energy cannot depend on which of the infinitely many equivalent unit cells represents a crystal and it must be extensive under replication.
These invariances and their first- and second-derivative consequences are exact statements whose residuals must vanish.
Their target is $0$, or, for the dimension of the Hessian null space, an exact integer $n$ ($n = 6$ for nonlinear molecules, $5$ for linear ones).
The improper part of the orthogonal group deserves emphasis: many equivariant architectures build in $\mathrm{SE}(3)$ but not $\mathrm{E}(3)$, so parity-controlled observables are structurally unprotected and must either be learned or be certified.

The second family is self-consistency and integrability.
A conservative model is generated by a single scalar potential $U(\mathbf{R})$: the energy, the forces $\mathbf{F}=-\nabla U$, the stress $\sigma = V^{-1}\partial U/\partial\varepsilon$ and the curvature $\mathbf{H}=\nabla\nabla U$ are not independent but must be mutually consistent.
Equivalently, the force field is a gradient field: its Jacobian (negative Hessian) is symmetric, work around a closed path vanishes and a symplectic integrator is exactly time-reversible.
On the extended configuration space of positions and cell strain the same requirement constrains all second-derivative blocks (position--position, strain--position and strain--strain) to be mutually symmetric, since all derive from one scalar.
Each consistency requirement is an identity with target $0$.
These probes are especially pointed for the growing class of models that predict forces or stresses directly rather than by differentiating an energy~\cite{darkside,orb,orbv3}.

The third family is statistical-mechanical equilibrium.
At equilibrium the model must obey the fluctuation relations of the ensemble its own dynamics generate.
The configurational temperature inferred from the spatial distribution must equal the kinetic temperature, the energy of every normal mode must satisfy equipartition and the Clausius virial of the sampled forces must balance the kinetic energy.
These are relations between independent channels (forces, curvature, positions and dynamics) that the training loss never forces to agree and their target is $1$.

The fourth family is regularity.
Independently of symmetry and self-consistency, a physical potential-energy surface is smooth: along a bond deformation its derivatives are bounded, it has no energy discontinuities and no spurious force reversals.
This probe, recently formalised as a bond-smoothness characterisation by Liu et al.~\cite{bsct} and tracked by community benchmarks~\cite{mliparena}, is well covered and we adopt a representative of it.

For a model that builds in exact symmetry and obtains forces by differentiating an energy, many of the symmetry and self-consistency residuals sit at the level of numerical noise by construction; there they act as correctness gates.
For so-called non-conservative models the respective probes describe if the large-scale pretrained models were able to learn these properties by training.

We now state each probe, the exact identity it certifies and precisely how its deviation is obtained from a model.
The technical details of every probe, including its equations and default parameters, are given in the Methods section.
Throughout, $\mathbf{R}$ denotes the $3N$ atomic coordinates, $E(\mathbf{R})$ the energy, $\mathbf{F}=-\partial E/\partial\mathbf{R}$ the forces and $\mathbf{H}=\partial^2 E/\partial\mathbf{R}^2$ the Hessian.

Each probe returns a non-negative deviation $d$ from its exact target and a normalised score:
\begin{equation}
    \mathcal{S}_\text{probe}=\exp \left(- \frac{d}{\lambda} \right),
    \label{eq:score_probe}
\end{equation}
where the scale $\lambda$ is fixed per probe and recorded with the result, so that a physically negligible deviation maps near $1$ and a clearly broken one near $0$.
A defect equal to $\lambda$ scores $1/e\approx0.37$, which makes $\lambda$ the interpretable sensitivity of each probe.
Most probes act on small test substrates.
These are the only system-specific inputs; because every model is interrogated on the same substrates and every target is exact, the certifications remain independent of chemistry.

\subsection*{Symmetry and invariance}

\subsubsection*{Rotational equivariance}
For any proper rotation the energy must be invariant and the forces must rotate rigidly with the structure.
This is the most basic geometric property of a potential energy surface, and architectures differ in whether they enforce it by construction or merely learn it from data.
We rotate a compact cluster about its centre of mass through a set of seeded random rotations and measure both the energy-invariance residual and the force-equivariance residual.
For an equivariant architecture both sit at the floating-point noise floor; for a model that learns orientation dependence from augmented data they do not, and the residual measures how much orientation the model has retained.
The rotations drawn here are proper, so the improper part of $\mathrm{O}(3)$ is certified separately by the stress selection rule that follows.

\subsubsection*{Improper-symmetry stress selection rule}
The exact surface is invariant under the full orthogonal group $\mathrm{O}(3)$, not only its proper part: reflections and inversion leave the energy unchanged, and by Neumann's principle the stress of a symmetric configuration must in addition be invariant under the configuration's own point group, which forbids specific components outright~\cite{nye}.
We exploit this to separate proper from improper symmetry using a single low-symmetry crystal in space group $Pmm2$, chosen so that one shear component of the stress is eliminated only by the mirrors, while two others are already forbidden by the proper twofold rotation and therefore act as built-in controls.
The first component is thus a pure improper-symmetry residual, structurally unprotected in an architecture that builds in $\mathrm{SE}(3)$ but not $\mathrm{E}(3)$, and the contrast between it and the controls localises a failure to the improper operations specifically.
A symmetry-forbidden shear stress exerts a systematic torque on the cell in variable-cell relaxations and NPT dynamics and slowly drives the angles away from their symmetric values.
It also comes with its own remedy, since averaging the stress over the structure and its parity image restores the forbidden zeros to machine precision at twice the inference cost, a special case of frame averaging~\cite{frameaveraging}.
The probe is skipped for models without a stress head.

\subsubsection*{Extensivity and representation invariance}
A periodic calculation refers to a crystal, not to the bookkeeping chosen to describe it, and three exact invariances follow.
First, the lattice basis is a gauge: any unimodular change of basis with the atoms rewrapped describes the identical crystal, so energy, Cartesian forces and stress must be unchanged.
Second, the description is extensive, so an $n$-fold supercell must return $n$ times the energy, replicated forces and the identical intensive stress; this is the size-consistency requirement familiar from quantum chemistry.
Third, the energy is invariant under a rigid translation of all atoms modulo the cell, a spurious dependence on the offset being the exact analogue of the egg-box effect that plagues grid-based electronic-structure codes~\cite{eggbox}, imported here as a certification.
Alongside these we test the remaining discrete invariance of this family, permutation of atom indices.
For strictly local models built on a radius graph all four residuals sit at machine precision; it can become discriminating for architectures with global pooling or all-to-all attention~\cite{allattention}, cell-conditioned features, or grid-based long-range modules, and the sheared representation additionally stresses neighbour-list construction in exactly the skewed cells that variable-cell relaxations and NPT trajectories visit.

\subsubsection*{Hessian null space and acoustic sum rule}
Relaxing a structure with the model itself yields a stationary point at which the Hessian is forced by symmetry to have a known null space, with eigenvectors fixed analytically by geometry: the rigid translations and the infinitesimal rotations must all be annihilated exactly.
For a periodic crystal the same statement at the zone centre is the acoustic sum rule.
We therefore relax the cluster to the model's own minimum, construct the null vectors directly from the relaxed geometry and measure how far the Hessian fails to annihilate them, alongside the count of near-zero eigenvalues, which must equal six for a generic cluster.
An eigenvalue that is appreciably negative at a relaxed minimum flags a spurious soft mode and lowers the score further.
Unlike phonon benchmarks, which score curvature against a reference, this is pure self-consistency with an exact target.

\subsubsection*{Many-body trimer probe}
A symmetric three-atom complex exposes behaviour that is invisible to pair scans, and we use it to certify two independent properties.
The first is the decay of the three-body term of the many-body expansion: as the apex atom recedes from the opposite edge, the genuine three-body energy must fall monotonically to zero, and we penalise both a finite residual at maximal separation and any non-monotonicity along the sweep.
The second is the symmetry of the force Jacobian: whenever forces derive from a scalar potential the Jacobian is minus the Hessian and the equality of mixed second derivatives makes it symmetric.
This is precisely the property a direct-force model need not possess, and any nonzero antisymmetric part certifies that the field cannot be written as the gradient of any potential; around a closed loop in configuration space such a field performs nonzero net work, so a potential energy is no longer single-valued and long trajectories drift in energy.
Testing this by assembling the Jacobian would cost $\mathcal{O}(N)$ force evaluations, which is what rendered the direct check impractical~\cite{pet_oam}; we avoid both by contracting the Jacobian with random direction pairs, which reduces the test to a scalar and leaves the cost independent of system size.
The qualifier transverse refers to those directions: a collinear bond scan probes only the longitudinal block, whereas random probe vectors span the full $3N$-dimensional space and reach the angular and off-diagonal non-conservativeness to which one-dimensional stretches are structurally blind.

\subsection*{Self-consistency and integrability}

\subsubsection*{Collinear conservativeness}
Along a one-dimensional bond stretch the force on the displaced atom, projected onto the bond axis, must equal minus the derivative of the energy with respect to that displacement.
Following the conservation-deviation diagnostic of MLIP-Arena~\cite{darkside,mliparena}, we displace one atom of a bonded pair along the bond axis over a symmetric scan, record the projected force and the energy at each point, and infer the energy-derived force from a high-order central difference.
The deviation is the mean absolute mismatch between the two over the interior of the scan, which vanishes exactly for a conservative model.
The high-order stencil keeps truncation error negligible, so a finite deviation reflects genuine energy/force inconsistency rather than discretisation.
Unlike the transverse facet of the trimer probe, this proxy stays strictly collinear and therefore catches only inconsistency with a longitudinal component.

\subsubsection*{Maxwell--Betti reciprocity}
The reciprocal theorem of linear elasticity states that the response at one degree of freedom to a unit displacement at another equals the reciprocal response; equivalently, the Hessian is symmetric.
We implement it as a finite-displacement response experiment: for a random pair of Cartesian degrees of freedom we push along the first and read the induced force at the second, then push along the second and read at the first, and measure the asymmetry between the two cross-responses.
Because the experiment evaluates off-diagonal Hessian elements directly in the coordinate basis, the deviation is near-zero by construction for any model that differentiates an energy, while for the unconstrained direct-force class the asymmetry is generically nonzero and the probe is discriminating.
The Maxwell--Betti probe is thus the canonical-basis, element-wise counterpart of the transverse trimer probe: both certify symmetry of the force Jacobian, the former by reading individual Hessian elements with unit coordinate pushes, the latter by contracting with random full-space vectors.

\subsubsection*{Strain--position reciprocity}
For a periodic system the natural configuration space is extended by the cell strain, and a conservative model is a single scalar $E(\mathbf{R},\boldsymbol{\varepsilon})$ on joint space.
Schwarz's theorem then constrains the mixed second-derivative block connecting the two sectors, so how the stress responds to an atomic displacement must equal, up to a sign, how the force responds to a strain.
A model whose force and stress heads disagree on it couples cell and internal degrees of freedom inconsistently, which is precisely the pathology that corrupts variable-cell relaxations and NPT dynamics.
At marginal extra cost we also complete the picture with the strain--strain block, whose Voigt symmetry tests the stress head's own integrability independently of the energy.
For a model whose stress is the exact strain derivative of the predicted energy both residuals vanish by construction and the probe is a gate; it is skipped when the model exposes no stress or the supplied cell is non-periodic.
Together with Maxwell--Betti reciprocity, this probe certifies all three blocks of the bordered second-derivative structure on positions and strain.

\subsubsection*{Stress--gradient consistency}
For a periodic cell the stress is the volume-normalised strain derivative of the energy.
We read the model's own stress once, then deform the cell along each of the six independent Voigt strains with the atoms scaled affinely and reconstruct the stress from central differences of the energy alone.
The deviation is the relative residual between the predicted and the reconstructed tensor.
Two model outputs are thereby cross-checked with no reference at all, which exposes inconsistent or approximate stress implementations.
The probe is skipped when the model exposes no stress (or the supplied cell is non-periodic).

\subsubsection*{Time-reversibility}
A constant-energy trajectory integrated forward, then backward after reversing the velocities, returns to its origin for any deterministic, position-dependent force propagated with a symplectic integrator.
We initialise seeded Maxwell--Boltzmann velocities on the cluster, propagate a fixed number of microcanonical steps, reverse all momenta, propagate the same number back, and report the per-atom root-mean-square return displacement.
The round-trip error is distinct from secular energy drift, which we record separately as the slope of the total energy along the trajectory: a model can have small per-step drift yet large irreversibility from hidden state or non-deterministic evaluation, or the reverse.

\subsection*{Statistical-mechanical equilibrium}

\subsubsection*{Configurational temperature}
A purely configurational thermometer relates the spatial fluctuations of the force to the mean curvature of the surface~\cite{rugh1997,butler1998}, and the temperature it reports coincides with the kinetic temperature only when the trajectory samples the Boltzmann distribution.
We relax to the model's own minimum, run a Langevin trajectory, and average both quantities over the sampled snapshots, estimating the curvature trace matrix-free by a Hutchinson average~\cite{hutchinson1989}.
The deviation is the departure of the ratio of the two temperatures from unity.
Being a finite-time average the ratio retains irreducible sampling noise and the exact target of unity holds only in the long-trajectory limit.
In general a non-conservative model samples no Boltzmann distribution at all, so the ratio drifts from unity.

\subsubsection*{Virial theorem}
The Clausius virial theorem is exact for any bound classical system in a stationary state, harmonic or not~\cite{clausius1870}: the virial of the internal forces balances twice the internal kinetic energy.
We evaluate it on the same Langevin trajectory produced for the configurational-temperature probe, forming the virial from the forces and the displacements from the model's own relaxed geometry, and the internal kinetic temperature after projecting out the rigid translation and rotation of the cluster, since the six unconfined degrees of freedom contribute kinetic energy but no confining virial.
The virial thermometer ties yet another independent pair of channels, positions and forces, to the kinetic energy.
Like the other equilibrium probes, the virial ratio is a finite-time average with irreducible sampling noise and its exact target holds in the long-trajectory limit.

\subsubsection*{Per-mode equipartition}
This probe rests on the classical equipartition theorem: in thermal equilibrium, every quadratic degree of freedom holds on average $k_BT/2$. 
Near a minimum, the potential-energy surface is harmonic to leading order and decomposes into independent normal modes, so each mode's stiffness times its mean-square thermal amplitude must equal $k_BT$. 
We test whether the model's own dynamics populate its normal modes as its own curvature demands, using two independent channels. 
The curvature channel diagonalises the Hessian at the relaxed minimum, yielding the normal modes and their frequencies $\omega_\alpha$. 
The dynamical channel reuses the Langevin trajectory of the configurational-temperature probe and projects each snapshot's mass-weighted displacement from the minimum structure onto these modes. 
This gives normal coordinates $Q_\alpha$, whose squares are averaged over the trajectory. 
The resulting per-mode ratio $\omega_\alpha^2\langle Q_\alpha^2\rangle/k_BT$ equals one for a self-consistent model and deviates mode by mode when curvature and dynamics disagree. 
Because each ratio retains irreducible finite-sampling scatter even for a perfect model, the scalar score instead measures how well the spectrum of the ratios centres on unity, using its median.

\subsection*{Regularity}

\subsubsection*{Bond-deformation smoothness}
Along a bond scan from compression through dissociation a physical surface is a smooth single well: no localised jumps or kinks in the force, a single zero crossing of the projected force, a single energy minimum and low force tortuosity~\cite{bsct}.
We scan the closest atom pair from moderate compression out into the dissociation tail, recording the projected force and the energy at each point.

\section*{Results}

\begin{figure*}[!t]
    \centering
    \includegraphics[width=0.85\textwidth]{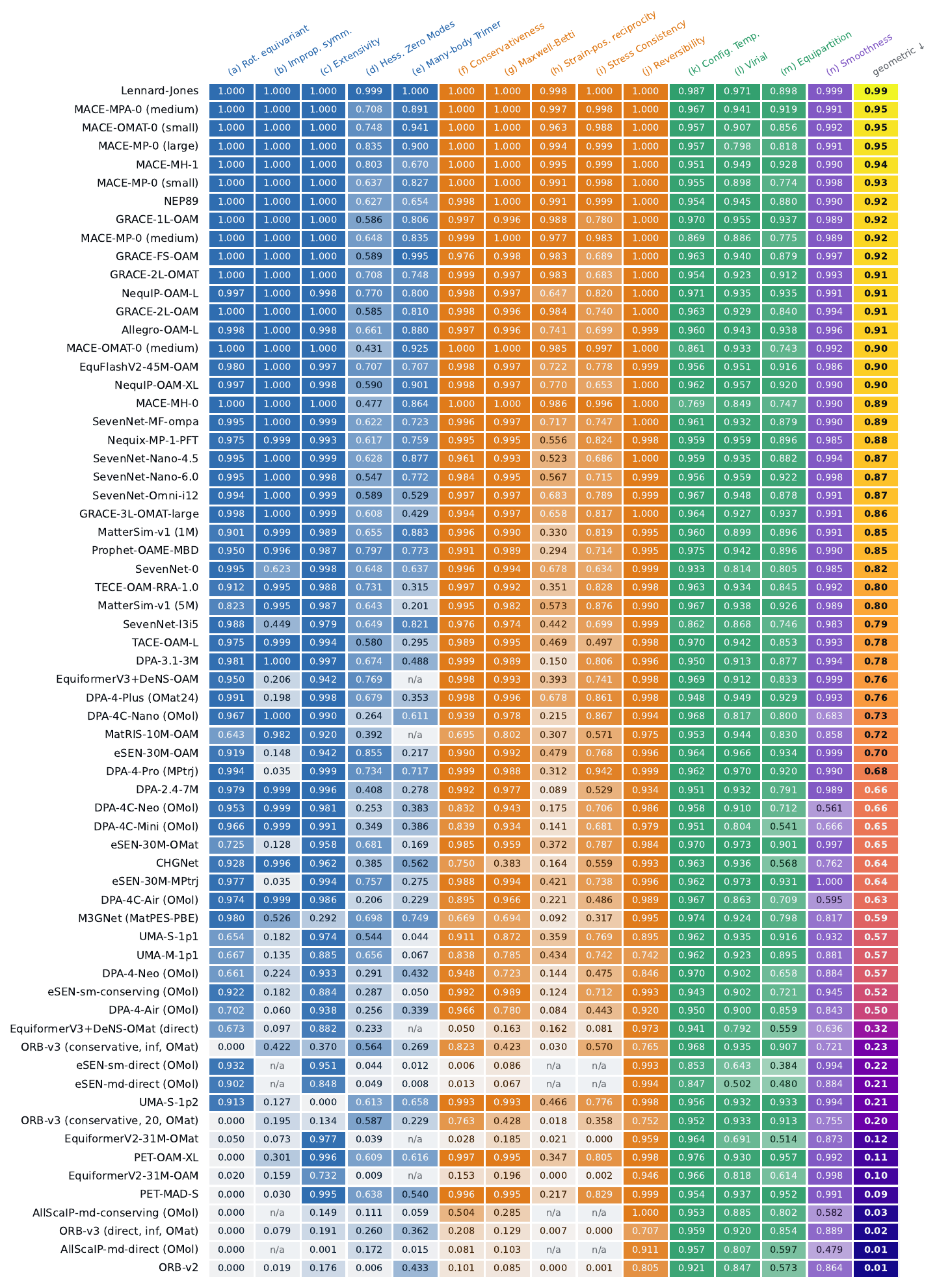}
    \caption{\textbf{Score matrix across models and probes.}
    Rows are models (with the Lennard-Jones potential as an analytical reference), columns the fourteen probes grouped by family; colour encodes the normalised score $\mathcal{S}\in[0,1]$ ($1=$ ideal).
    Probes that do not apply to a model are marked `n/a' and excluded from the geometric-mean aggregate of the rightmost column.}
    \label{fig:heatmap}
\end{figure*}

We apply the certification suite to two sets of models.
The first is a set of $64$~contemporary large-scale pretrained potentials spanning equivariant message-passing, transformer and descriptor-based architectures, energy-conserving and direct-force heads and $\mathrm{E}(3)$- as well as $\mathrm{SE}(3)$-equivariant designs.
Every model is interrogated on the identical chemistry-independent substrates, built for the same ten elements (Li, Be, B, C, Na, Mg, Al, Si, P and S) from the same seed, so that no model sees a structure tailored to it; each entry of Figure~\ref{fig:heatmap} is the mean over the ten elements (Methods).
A complete run consumes $3.5$--$4.2\times10^{4}$ single-point evaluations on the small systems and between five minutes and three hours (CPU bound models) of wall time on a single consumer workstation (median $24$~minutes for all ten elements, i.e.\ of order two minutes per element).
Figure~\ref{fig:heatmap} collects all fourteen probe scores and the geometric-mean aggregate for each model.
The second set is a series in which the \textsc{MACE-MP-0} (small) universal model is fine-tuned on one and the same system by three different methodologies, which isolates the effect of the fine-tuning protocol on the reference-free scores (Figure~\ref{fig:ft_study} and Supplementary Note~17).

\subsection*{The analytic control sets the floor}
Alongside the machine-learning models we score the analytic Lennard-Jones potential on the identical substrates and with the identical settings; the Supplementary Information (Supplementary Notes~1--16) additionally validates every probe against analytic modulations designed to defeat it.
The Lennard-Jones reference passes every identity-based probe at the numerical floor, with raw residuals between $10^{-16}$ and $10^{-5}$ in the probe's own units.
Its three trajectory-based equilibrium scores ($0.987$, $0.971$ and $0.898$ for configurational temperature, virial and equipartition, respectively) and hence its aggregate of $0.99$ measure the finite-sampling floor of the default trajectory lengths, not a defect of the potential.
Scores in that range are therefore to be read as at the floor rather than as slightly imperfect, and the control row of Figure~\ref{fig:heatmap} is the scale against which every other row should be compared.

\subsection*{Certified quality spans two orders of magnitude at nominally equal accuracy}
All $64$~pretrained models report near-ab initio energy and force errors on held-out data~\cite{mliparena,matbench}, yet their aggregates span from $0.95$ down to $0.01$.
At the top, \textsc{MACE-MPA-0} (medium, $0.95$) is statistically indistinguishable from most of the other \textsc{MACE} checkpoints, the \textsc{GRACE} family (besides the $3$L version) and the \textsc{NEP89} universal model ($0.91$--$0.95$).
These are followed by \textsc{NequIP}, \textsc{Allegro}, \textsc{EquFlashV2}, \textsc{SevenNet}, \textsc{Nequix} and \textsc{MatterSim} ($0.80$--$0.91$).
A broad middle field, from $0.50$ to $0.80$, contains models of the \textsc{TACE}, \textsc{DPA}, \textsc{EquiformerV3}, \textsc{MatRIS}, \textsc{eSEN} and \textsc{UMA} families alongside the earlier \textsc{M3GNet} and \textsc{CHGNet} potentials.
At or below $0.32$ sit for example the direct-force and unconstrained-backbone models: \textsc{EquiformerV3+DeNS-OMat (direct)} ($0.32$), the \textsc{ORB-v3} variants ($0.02$--$0.23$), \textsc{UMA-S-1p2} ($0.21$), the \textsc{EquiformerV2} checkpoints ($0.10$ and $0.12$), the \textsc{PET} models ($0.09$--$0.11$) and the \textsc{AllScaIP} variants ($0.01$--$0.03$), with \textsc{ORB-v2} closing the field at $\mathcal{S}_\mathrm{overall}=0.01$.

The aggregate is deliberately a geometric mean, and the difference from an arithmetic one is not cosmetic.
Twelve models differ by more than $0.3$ between the two conventions: \textsc{PET-OAM-XL} scores $0.75$ arithmetically but $0.11$ geometrically, \textsc{PET-MAD-S} $0.72$ against $0.09$, \textsc{UMA-S-1p2} $0.74$ against $0.21$, \textsc{eSEN-sm-direct (OMol)} $0.54$ against $0.22$, and the \textsc{ORB}, \textsc{EquiformerV2}, \textsc{AllScaIP} models.
The \textsc{PET} models and \textsc{UMA-S-1p2} in particular sit near the floor on twelve or thirteen of the fourteen probes and fail one or two outright, which is precisely the profile an average conceals and a production trajectory discovers.
Reporting the axes separately, as in Figure~\ref{fig:heatmap}, therefore carries information that no scalar can.

Read across the columns, the suite separates the equivariant from the non-equivariant, the conservative from the direct and the $\mathrm{E}(3)$- from the $\mathrm{SE}(3)$-equivariant architectures.
That this separation appears at all is expected; what the certification adds is that it is now quantitative and measured on a common absolute scale, so the question of how much of the missing physics a large training set can supply in place of an architectural constraint becomes an empirical one.
Several direct-force models, for example, score markedly better than their class on the conservativeness probe, which is a statement about what training has achieved rather than about what the architecture guarantees~\cite{darkside}.
We therefore do not read these results as an argument for excluding any architecture class.
Direct force prediction avoids the backward pass and is correspondingly faster per step, which can make it the rational choice for high-throughput screening or very long trajectories, particularly in combination with schemes in the spirit of Tuckerman et al., who showed in 1992 that periodic corrections to an approximate force can keep a trajectory stable and physical without paying for an exact evaluation at every step~\cite{tuckerman1992}.
The purpose of the suite is to make the trade-off explicit rather than to settle it.

Two limitations should be kept in view when reading Figure~\ref{fig:heatmap}.
First, precision and architecture are confounded in the present model set: the $14$ potentials that expose a \texttt{float64} path all belong to the \textsc{MACE} and \textsc{DPA} families and have a median aggregate of $0.91$, against $0.71$ for the $50$ \texttt{float32} models.
Because the suite chooses finite-difference steps from the model's declared working precision (Methods) and normalises the symmetry residuals at $10^{-3}$, a \texttt{float32} model's residual contains an arithmetic contribution that a \texttt{float64} model does not, so its scores on the sharpest probes are conservative lower bounds.
Second, five models (the \textsc{EquiformerV2}, \textsc{EquiformerV3} and \textsc{MatRIS} checkpoints) cannot evaluate an isolated atom at all and therefore cannot supply the monomer terms of the many-body expansion; their trimer probe is recorded as not applicable rather than as a failure, although the inability itself is a practical restriction worth noting.

\subsection*{Symmetry is either built in or measurably imperfect}
The equivariance column splits the field cleanly.
Architectures with built-in rotational equivariance, the \textsc{MACE}, \textsc{GRACE}, \textsc{NEP}, \textsc{NequIP}, \textsc{Allegro} and \textsc{SevenNet} families as well as \textsc{DPA-2}, \textsc{DPA-3} and \textsc{DPA-4-Pro}, return $\geq0.98$ and mostly $1.000$ to the numerical floor, whereas the unconstrained backbones that must learn symmetry from data or augmentation, the \textsc{PET}, \textsc{ORB} and \textsc{AllScaIP} models, return scores near $0.000$.
Between these two regimes the residual behaves as a continuous quantity rather than a binary label: \textsc{MatterSim}, which uses invariant descriptors rather than steerable tensor features, reaches $0.90$ (1M) and $0.82$ (5M), and the \textsc{UMA} and OMol-trained \textsc{DPA-4} checkpoints occupy $0.65$--$0.91$.
A practical consequence for the unconstrained backbones is that the remedy is known and cheap: enlarging the training set with rotated copies of existing structures targets exactly the residual this column measures, and the column provides the metric by which such an augmentation can be verified rather than assumed.

The improper part of $\mathrm{O}(3)$ behaves exactly as the taxonomy anticipates, and is the most sharply bimodal column of the suite.
The parity-odd stress remainder stays at machine precision for the \textsc{MACE}, \textsc{GRACE}, \textsc{NEP} and \textsc{MatterSim} families as well as for \textsc{DPA-2} and \textsc{DPA-3}, but is large for the \textsc{UMA} models ($0.13$--$0.18$), \textsc{PET-MAD-S} ($0.03$), \textsc{eSEN-30M-MPtrj} ($0.04$), \textsc{DPA-4-Pro} ($0.04$), \textsc{DPA-4-Neo} ($0.22$), \textsc{eSEN-30M-OAM} ($0.15$) and \textsc{PET-OAM-XL} ($0.30$).
The affected models are not respecting parity information by construction and are only able to learn it by data, which buys efficiency by restricting the enforced symmetry to the proper subgroup ($\mathrm{SO}(3)$).
That the same construction yields both the machine-precision and the failing rows shows that pretraining at scale can instil parity symmetry where the architecture does not enforce it, but does not have to; it is not something a training-set size alone predicts.
The failure is also localised rather than diffuse: because the probe carries its own proper-rotation controls in $\sigma_{xz}$ and $\sigma_{yz}$, a low score is attributable to the improper operations specifically.
Where it occurs, the remedy of Equation~\eqref{eq:paritysym} is available at twice the inference cost, by evaluating the model on the structure and on its parity image and averaging the two stresses.

The extensivity and representation column is the one on which formally local architectures can nonetheless be caught, because it tests the implementation as much as the functional form.
The \textsc{MACE} and \textsc{GRACE} models return residuals at machine precision, as a strictly local radius-graph model must.
By contrast the transformer-style backbones that connect every atom in the cell are exposed by the supercell comparison: \textsc{AllScaIP-md-conserving} (OMol) reaches $0.15$ and \textsc{AllScaIP-md-direct} $0.001$.
\textsc{M3GNet} scores $0.29$ despite a strictly local three-body construction, which points at the neighbour-list handling in the sheared unimodular cell rather than at the model: exactly the skewed cells that variable-cell relaxations and NPT trajectories visit.

The many-body trimer probe is the harshest column of the symmetry family, with a median of $0.61$ and only four models above $0.9$.
\textsc{GRACE-FS-OAM} ($0.99$), \textsc{MACE-OMAT-0} (small, $0.94$; medium, $0.93$) and \textsc{NequIP-OAM-XL} ($0.90$) lead; at the other end the \textsc{eSEN} and \textsc{AllScaIP} OMol checkpoints and \textsc{UMA-S-1p1} fall to $0.01$--$0.06$, meaning that their three-body term does not decay as the apex recedes, so a spurious many-body interaction survives at separations at which the physical one has long vanished.

\subsection*{Non-conservative force fields fail the entire integrability family coherently}
The direct \textsc{ORB}, \textsc{eSEN}, \textsc{EquiformerV2}, \textsc{EquiformerV3} and \textsc{AllScaIP} models score at or near zero simultaneously on collinear conservativeness ($\leq0.21$), Maxwell--Betti reciprocity ($\leq0.20$), stress--gradient consistency ($\leq0.09$) and strain--position reciprocity ($\leq0.17$).
This is the signature of forces that are not the gradient of any scalar and of stresses that are not the strain derivative of the predicted energy, and its coherence across four independent probes is itself evidence that the probes measure what they claim to: a model that merely had noisy outputs would not fail all four in the same direction.
The failure is also graded rather than absolute.
\textsc{ORB-v3 (direct, `inf', OMat)} is the strongest of the direct class on the conservativeness column, reaching $0.21$ where the direct \textsc{eSEN} checkpoints reach $0.01$, so a direct-force model can evidently learn part of the integrability it does not enforce, without reaching the machine-precision regime that differentiating an energy delivers for free.
Within the \textsc{ORB-v3} family the comparison is particularly clean, since the conservative and direct heads share a backbone: conservativeness rises from $0.21$ to $0.82$ and Maxwell--Betti from $0.13$ to $0.42$ when the gradient head is used, which quantifies what the architectural choice buys on a common scale.

The same models pass time-reversibility at scores $\geq0.70$, and indeed the reversibility column is one of the least discriminating of the suite (interquartile range $0.02$).
This is by design rather than a shortcoming: velocity Verlet is exactly reversible for any deterministic, position-dependent force, conservative or not, so the round trip isolates hidden state and non-deterministic evaluation specifically and is silent on non-conservativeness, which the other four columns of this family address.

\subsection*{First-derivative consistency does not imply second-derivative consistency}
Among the energy-conserving models the stress--gradient probe is passed almost universally, the \textsc{UMA} family for instance reaching $\geq0.74$, yet the strain--position reciprocity behind it fails widely, for the same \textsc{UMA} models at $\leq0.47$.
Only the \textsc{MACE}, \textsc{GRACE} and \textsc{NEP89} models keep the mixed block consistent ($\geq0.96$, with \textsc{GRACE-3L-OMAT-large} the single exception at $0.66$).
A stress head can therefore agree with the energy along the six Voigt directions while disagreeing with the force head about the internal-strain tensor $\Lambda$: the two heads are each integrable in their own sector and mutually inconsistent across sectors.
Because $\Lambda$ is precisely the quantity that couples internal coordinates to cell strain, this is the block whose corruption shows up as drifting cell angles in variable-cell relaxations and NPT dynamics rather than as a large energy error, and, after the parity rule, it is the most discriminating column of the suite (interquartile range $0.56$).

\subsection*{No pretrained model reproduces the curvature at its own minimum}
The Hessian null-space probe is the only column on which not a single machine-learning model reaches the analytic floor: the control scores $0.999$, the best models reach $0.86$ (\textsc{eSEN-30M-OAM}) and $0.84$ (\textsc{MACE-MP-0}, large), the median is $0.61$ and the weakest models (\textsc{ORB-v2}, the \textsc{EquiformerV2} checkpoints, the direct \textsc{eSEN} variants) fall to $0.01$--$0.05$.
The probe is evaluated at each model's own relaxed geometry, so it is not a statement about disagreement with a reference: it says that at a configuration the model itself declares stationary, the rigid-body modes that must be exactly flat carry finite curvature, and in the weakest cases the spectrum additionally contains spurious imaginary modes.
Both defects are fatal for phonons, for harmonic free energies and for any transition-state search that relies on the curvature.
We therefore recommend that harmonic analyses built on machine-learned potentials be validated deliberately rather than assumed, and that this column in particular be consulted before a model is used for vibrational properties.

\subsection*{The equilibrium family is a floor test, not a ranking}
The three statistical-mechanical probes behave differently from the identity-based ones, and the distinction matters for how they should be read.
Across the $64$~models the configurational temperature has a median of $0.96$ and an interquartile range of $0.01$, the virial ratio $0.93$ and $0.05$, and the per-mode equipartition $0.87$ and $0.15$, all sitting essentially near the Lennard-Jones control values of $0.987$, $0.971$ and $0.898$.
At the default trajectory lengths these probes are therefore dominated by finite-sampling noise and cannot separate well-behaved models from one another; they function as gates against gross violation, and resolving them into a ranking axis would require substantially longer trajectories.

What they do resolve is the direct-force class, which holds the four lowest virial scores, although the conservative \textsc{DPA-4C-Mini} and \textsc{CHGNet} reach similarly low equipartition scores ($0.54$ and $0.57$).
The clear outliers are \textsc{eSEN-sm-direct} ($0.38$ equipartition, $0.64$ virial), \textsc{eSEN-md-direct} ($0.48$, $0.50$), \textsc{EquiformerV2-31M-OMat} ($0.51$, $0.69$) and \textsc{EquiformerV3+DeNS-OMat (direct)} ($0.56$, $0.79$).
The mechanism is the one the identities predict: a force field that is not a gradient field generates dynamics that samples no Boltzmann distribution, so the amplitude with which a mode is populated stops matching the stiffness measured at the minimum, and the sampled state is not stationary in the sense the virial theorem requires.
These probes thus provide an independent, dynamics-based confirmation of the static integrability failures of the previous section, obtained from a completely different functional of the same model.

\subsection*{Fine-tuning can degrade certified quality while every reported error improves}
\begin{figure}[t]
    \centering
    \includegraphics[width=\columnwidth]{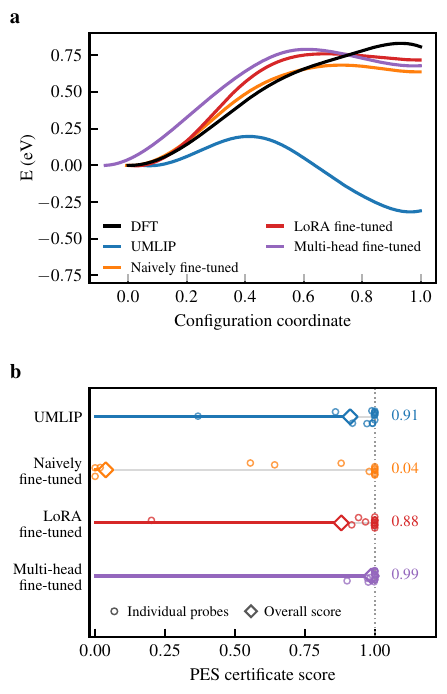}
    \caption{\textbf{Certification of fine-tuned models.}
    \textbf{a}~Energy along the configuration coordinate of the sulfur-vacancy jump in MoS$_2$, obtained with the universal \textsc{MACE-MP-0} (small) model and with three fine-tuned versions, against the density-functional reference.
    All three fine-tuned models reproduce the reference barrier.
    \textbf{b}~The corresponding reference-free certification: individual probe scores (dots) and the geometric-mean aggregate (bar) for each model.
    The naively fine-tuned model, which attains the lowest held-out force error of the series, is the only one whose certificate collapses, and the only one that fails to sustain molecular dynamics on a related sulfur crystal.}
    \label{fig:ft_study}
\end{figure}

Figure~\ref{fig:ft_study} tracks \textsc{MACE-MP-0} (small) fine-tuned on the two-dimensional layered molybdenum disulfide system with multiple sulfur vacancies~\cite{li2018,spetzler2024}.
The sulfur-vacancy jump is a demanding target: the barrier is high and the relevant configurations are off-equilibrium, so they are largely absent from universal training sets and the corresponding potential-energy surface is predicted poorly by state-of-the-art universal models~\cite{flototto2026large,atk_found}.
The fine-tuning dataset comprises $572$~first-principles data points distributed roughly uniformly over $30$~bins along the configuration coordinate of the jump (minimum $14$ points per bin).
We fine-tune by three techniques: naive fine-tuning, in which all weights are free; LoRA fine-tuning, in which small additional low-rank matrices are trained while the original weights stay frozen; and multi-head replay fine-tuning, in which two heads are trained simultaneously, one on a subset of the universal training data and one on the fine-tuning set~\cite{tompa2026fine,atk_size}.

All three fine-tuned models reproduce the target property.
Figure~\ref{fig:ft_study}a shows the predicted jump barriers against the density-functional reference, and all three track it far better than the universal model does.
The reported errors agree: on a held-out set of $200$~configurations the naive model reaches the lowest force error of the three ($0.055$~eV/{\AA}, against $0.060$ for LoRA and $0.151$ for multi-head), and all three improve on the universal model (Supplementary Note~17).
By the two criteria a practitioner would normally apply, the physical observable and the held-out error, the naively fine-tuned model is the best of the series.

The certification says the opposite.
The aggregate scores are $0.91$ for the universal model, $0.88$ for LoRA and $0.99$ for multi-head, but $0.04$ for the naive model (Figure~\ref{fig:ft_study}b; per-probe scores in Supplementary Note~17).
The collapse is not diffuse: the naive model still passes equivariance, parity, representation invariance, conservativeness and reversibility at machine precision, and fails on the many-body trimer term (score below the $10^{-12}$ clipping floor of Equation~\eqref{eq:overall}), strain--position reciprocity, the Hessian null space ($0.02$), equipartition ($0.56$) and smoothness ($0.64$).
In other words, unconstrained fine-tuning on a narrow dataset left the built-in symmetries intact, as it must, and destroyed the transferable parts of the surface that the narrow dataset does not constrain.

The consequence is immediate in simulation.
Running the four models on octathiocane~\cite{zeng2021}, a crystal built entirely of S$_8$ rings and thus chemically adjacent to the fine-tuning system but not part of it, the universal, LoRA and multi-head models complete $10{,}000$ steps of Nos{\'e}--Hoover-chain dynamics at $300$~K without incident, while the naive model heats to nearly $2{,}000$~K within the first $100$~steps and returns non-finite coordinates at step~$140$.
The certification predicted this from the probes alone, at a cost of a few thousand single-point evaluations and without any reference calculation, whereas the held-out errors and the target barrier did not.
The series also shows that the certificate is not simply kind to parameter-efficient methods: the LoRA model records the lowest smoothness score of the four ($0.20$), localising a roughness in the bond scan that is real but, on this system, not severe enough to destabilise the trajectory.

\section*{Discussion}

Two results here are worth separating from the details.
The first is that reference-free identities have enough resolving power to be useful: across $64$~potentials whose reported accuracies are broadly comparable, the certified quality spans two orders of magnitude, and the spread is structured rather than noisy, tracking architectural commitments that are otherwise only describable as labels.
The second is that the resolution is orthogonal to accuracy rather than redundant with it.
In the fine-tuning series the ordering by held-out force error and the ordering by certification are effectively inverted, and the certification is the one that predicts which model survives a trajectory.
A metric that agrees with accuracy on the easy cases and disagrees on exactly the cases that matter is the kind of complement worth adding to a validation protocol.

The probes are deliberately complementary and their coverage has structure.
A collinear conservativeness scan is, by construction, insensitive to a purely transverse non-conservative force, which is exactly why the transverse trimer probe exists; a symplectic round trip is reversible in exact arithmetic for any deterministic, position-dependent force, so it isolates hidden state and non-deterministic evaluation rather than non-conservativeness; the three equilibrium probes triangulate the same trajectory through independent functionals; and the parity probe carries its own proper-rotation control, so a failure localises to the improper operations specifically.
Reporting the axes separately therefore conveys where a model fails, not merely that it does, which is what turns a score into an actionable diagnosis.

In the broader picture, the conservativeness axis is also where our results can be checked most directly against independent evidence, and they agree with it.
Recent studies running molecular dynamics with direct-force potentials, \textsc{ORB-v2} and the direct \textsc{ORB-v3} variants among them, report a systematic drift of the total energy with simulation time in microcanonical runs~\cite{darkside,simple2025transformer,orbv3}, and, even under a thermostat, a centre-of-mass drift some six orders of magnitude larger than for gradient-based models~\cite{mliparena}.
Our suite singles out the same models as non-conservative, but reaches that verdict from static scans in minutes on commodity hardware, rather than from long and costly trajectories in which the defect only becomes visible once it has accumulated.
That a direct-force architecture cannot conserve energy exactly is, of course, expected from its construction.
What the architecture does not fix is how closely the trained model approximates a conservative field: unconstrained models can differ markedly in how well they conserve energy in practice~\cite{darkside,simple2025transformer}, and conservativeness has been found hard to learn and hard to monitor~\cite{darkside}.
It is this learned degree of conservativeness that the probes quantify.
The same comparison shows what the separation of axes adds.
In the \textsc{ORB-v3} microcanonical tests, a model whose forces are obtained as energy gradients still drifts once a neighbour limit is imposed, because the limit introduces small discontinuities into the surface~\cite{orbv3}.
Energy drift alone cannot tell these two defects apart, whereas our suite reports non-conservativeness on its own axis, through the conservativeness scans.

What the suite does not do is equally important.
A high certificate is necessary, not sufficient: the analytic Lennard-Jones control scores~$0.99$ and is nevertheless a poor description of any real material, because internal consistency says nothing about whether a model has learned the right chemistry.
None of these probes replaces accuracy benchmarks; they are complementary axes that error metrics provably cannot capture, and they are most informative when reported alongside them.
The converse property is what makes them worth adding: because the targets are exact and the substrates are generated on demand from a seed rather than drawn from a fixed set, there is no test set to overfit and no reference calculation whose functional or basis set could be contested.
Optimising a model against these identities means making it more nearly a potential-energy surface, which is the thing we wanted in the first place.
The limitations are correspondingly concrete, and we have marked them where they arise: the equilibrium family is sampling-limited at the default trajectory lengths, the sharpest symmetry residuals are precision-limited for \texttt{float32} models, and the substrates are small and elemental by design, so the suite certifies the surface a model produces rather than the chemistry it covers.

The fine-tuning series turns the litmus test of the introduction into a workflow recommendation.
Fine-tuning on a narrow dataset is precisely the setting in which held-out errors look best, since the test set is usually drawn from the same narrow distribution, while the quality of the surface away from that distribution can deteriorate unobserved.
Our series shows this is not hypothetical: the same base model, the same data and three different protocols produce certificates from $0.04$ to $0.99$, and the protocol that scores best by force error scores worst by certification.
How strongly a model degrades under narrow fine-tuning also depends on the architecture: in our earlier fine-tuning studies, \textsc{ORB-v2} models fell behind \textsc{MACE} and \textsc{GRACE}, most visibly for the smallest training sets~\cite{atk_found,atk_size}, in line with the certificates reported here.
Because the probes need no reference data, run in minutes on commodity hardware and return absolute scores, they can be evaluated at every fine-tuning checkpoint as a physics regression test, catching such degradation when it begins rather than when a production trajectory fails.

The design philosophy has a precedent in density-functional theory.
There, the vocabulary that made functionals comparable across constructions was not their parametrisation but the list of exact constraints on the universal functional that each one satisfies: PBE was built to satisfy a handful of them, SCAN all seventeen known to apply at its rung, and constraint satisfaction became the architecture-agnostic currency of functional quality~\cite{pbe,scan}.
The identities collected here are intended to play the same role for machine-learning interatomic potentials.
A model may satisfy an identity by construction, through an equivariant architecture or forces obtained by differentiating an energy, or approximately by learning; the certification measures the satisfaction and is indifferent to the route.
We regard this as a more informative description of a model than an architectural label: whether a network is $\mathrm{E}(3)$- or $\mathrm{SE}(3)$-equivariant is a design decision, whereas whether the symmetry-forbidden shear stress vanishes on a $Pmm2$ crystal is a measurable fact, and the parity column of Figure~\ref{fig:heatmap} shows that the two do not always coincide.
Where a constraint is violated, restoration is sometimes cheap: the parity average of Equation~\eqref{eq:paritysym} recovers the improper selection rules at twice the inference cost, mirroring the constraint-restoration strategies of functional design.

Machine-learned potentials are now used where the quantity of interest is not an energy but a trajectory, a phonon spectrum or a rate, and for those the relevant question is not how closely a model reproduced a set of snapshots but whether it is a potential-energy surface at all.
Asking that question does not require a reference calculation, only the identities the exact surface has to satisfy.
We expect the list assembled here to grow, and we have built the implementation so that adding an identity means adding one routine with the same result object.
A certificate of this kind will not tell anyone which model is most accurate; it will tell them which models can be trusted to behave like physics when taken outside the distribution they were fitted on, which is where they are increasingly asked to work.

\section*{Methods}

All probes interact with a model only through its energy, forces and (where available) stress, mediated by a thin engine that counts every single-point evaluation, so each probe reports and budgets its cost; one forward pass that returns energy and forces from a single geometry counts once and optimisers and integrators are driven through the same path so that relaxation and dynamics are counted automatically.

Second derivatives are never taken inside the model: the Hessian and Jacobian-vector products are obtained matrix-free from central differences of forces (second-order central; the resulting discretisation floor is quantified in Supplementary Note~1), so all operations depend only on $E$, $\mathbf{F}$ and the stress $\sigma$ as exposed by the potential.
The Hessian column $j$ is assembled as
\begin{equation}
  \mathbf{H}_{:,j}
  = -\frac{\mathbf{F}(\mathbf{R}+\epsilon\,\mathbf{e}_j)-\mathbf{F}(\mathbf{R}-\epsilon\,\mathbf{e}_j)}{2\epsilon},
  \label{eq:hessfd}
\end{equation}
where $\mathbf{e}_j$ displaces the $j$-th Cartesian degree of freedom and $\epsilon=10^{-3}\,\text{\AA}$ by default; the full $3N\times 3N$ matrix therefore costs $2\cdot 3N=6N$ force evaluations and we symmetrise $\mathbf{H}\leftarrow\tfrac12(\mathbf{H}+\mathbf{H}^{\!\top})$ since the finite-difference operator is symmetric only to discretisation order.
A Hessian--vector product along an arbitrary direction $\mathbf{v}$ needs only two evaluations, independent of system size,
\begin{equation}
  \mathbf{H}\,\mathbf{v}
  = -\,\lVert\mathbf{v}\rVert\,
  \frac{\mathbf{F}(\mathbf{R}+\epsilon\,\hat{\mathbf{v}})-\mathbf{F}(\mathbf{R}-\epsilon\,\hat{\mathbf{v}})}{2\epsilon},
  \quad
  \hat{\mathbf{v}}=\mathbf{v}/\lVert\mathbf{v}\rVert,
  \label{eq:hvp}
\end{equation}
with the Jacobian--vector product $\mathsf{J}\mathbf{v}=\partial\mathbf{F}/\partial\mathbf{R}\ \mathbf{v}=-\mathbf{H}\mathbf{v}$ obtained from the same two forces.

Units are those of the atomic-simulation environment package (\textsc{ASE}) throughout: energy in eV, length in {\AA}, forces in eV/{\AA}, stress in eV/{\AA$^3$} and temperature in K~\cite{ase}.
Hessians are assembled column-by-column from central differences of forces, costing $6N$ force evaluations; Hessian- and Jacobian-vector products use the same central difference along a single direction at two evaluations, so curvature-based probes remain matrix-free and architecture-independent.
Stationary points used by the null-space, strain--position, configurational-temperature, virial and equipartition probes are obtained by relaxing with the model's own forces using a standard optimiser implemented in \textsc{ASE} to a force tolerance (default setup: BFGS optimizer with $F_{max} = 10^{-3}$~eV/{\AA}), returning the model's stationary point; no reference is involved at any stage.
The strain–position probe relaxes positions and cell jointly (Fr\'echet cell filter, $F_{\max}=10^{-2}$~eV/{\AA}, at most $200$~steps).
Finite-difference steps use central stencils, with a small step chosen to straddle floating-point noise and anharmonicity for the curvature probes; near-zero eigenvalue counts use a tolerance relative to $\lVert\mathbf{H}\rVert$ and frequencies are mass-weighted.

The identities are universal; only the test substrate is supplied per model.
By default we build small, in-domain structures from the model's element(s): a compact $13$-atom cluster, a symmetric mirror-related trimer, a small periodic crystal and a low-symmetry $Pmm2$ crystal for the improper-symmetry probe, with all nearest-neighbour distances taken from covalent radii, so that universal and narrowly fine-tuned models are tested on equal footing with no system-specific knowledge; the user may instead pass any structure of their own.
Every randomised probe (random rotations, Hutchinson sign vectors, initial velocities, response-pair selection) is seeded, so a given seed reproduces the reported numbers and every probe tracks the number of model evaluations it consumes so that the whole suite stays well within a budget of order $10^3$ single-point calls per substrate; a per-probe call budget can be imposed, in which case probes shrink their sampling rather than overrun it.

Three probes (improper-symmetry stress, stress--gradient consistency and strain--position reciprocity) require a stress tensor.
Models that expose none, here the two direct \textsc{eSEN} and both \textsc{AllScaIP} checkpoints trained on \textsc{OMol25}, have these probes recorded as not applicable rather than as failures, and their aggregate is the geometric mean over the eleven remaining probes.
The many-body trimer probe is likewise recorded as not applicable for the five models that cannot evaluate an isolated atom and therefore cannot supply the monomer terms of the many-body expansion (the \textsc{EquiformerV2}, \textsc{EquiformerV3} and \textsc{MatRIS} checkpoints).
Aggregates are therefore taken over different numbers of probes for different models, and per-probe scores rather than the aggregate should be used when comparing a stress-free model with a stress-providing one.

Every finite-difference probe fixes its step from the model's declared working precision: the step is bounded below by the smallest spacing at which the model's own round-off does not dominate the difference. 
Where a probe's nominal grid would fall below that bound, the grid is thinned rather than the range shrunk, the scan range is a physical choice, and the run is marked as resolution-limited. 
For a \texttt{float32} model, force differences are taken with a step of $\epsilon=0.0106$~{\AA} rather than $0.0010$~{\AA}, and the collinear conservativeness scan is reduced from $31$~points spaced $0.0100$~{\AA} apart to $9$~points spaced $0.0375$~{\AA} apart.
A residual reported for a \texttt{float32} model is therefore an upper bound, part of it is truncation at the coarser step.
On the analytic Lennard-Jones control, where the entire residual is discretisation error by construction, that floor is $6.0\times10^{-4}$~eV/{\AA} for silicon against $3.3\times10^{-6}$eV/{\AA} at \texttt{float64} (Supplementary Note~1).

The suite is implemented as an open-source Python package, \textsc{pescert}, that wraps any potential exposed as an \textsc{ASE} calculator or as raw energy/force callables and exposes each probe behind a uniform interface returning the exact target, the deviation, the normalised score, the call count and per-mode diagnostics.
The package, its analytic-potential test suite, the demonstration script that produces the validation figures and the scripts used in this study are publicly available.

\subsection*{Probe specifications}

This subsection specifies each probe in full: the identity it certifies, the substrate, the sampling and finite-difference scheme, the deviation and the scale $\lambda$ entering Equation~\eqref{eq:score_probe}.
The probes appear in the order of the Certifications section.

\subsubsection*{Rotational equivariance}
This probe certifies the invariance of the energy and the equivariance of the forces under proper rotations.
For any rotation $\mathsf{Q}\in\mathrm{SO}(3)$ the energy is invariant and the forces rotate with the structure,
\begin{equation}
  E(\mathsf{Q}\mathbf{R})=E(\mathbf{R}),
  \qquad
  \mathbf{F}_i(\mathsf{Q}\mathbf{R})=\mathsf{Q}\,\mathbf{F}_i(\mathbf{R}).
  \label{eq:equivariance}
\end{equation}
The substrate is a compact cluster of an element known to the model (the $13$-atom cluster cut from the element's experimental reference crystal).
We evaluate $E(\mathbf{R})$ and $\mathbf{F}(\mathbf{R})$ once, draw a set of $n_{\mathrm{rot}}=8$ proper rotations from a seeded random stream and for each rotate the cluster rigidly about its centre of mass, $\mathbf{R}\mapsto\mathsf{Q}(\mathbf{R}-\mathbf{R}_{\mathrm{cm}})+\mathbf{R}_{\mathrm{cm}}$, re-evaluating energy and forces.
The deviation averages the energy-invariance residual $d_E=\lvert E(\mathsf{Q}\mathbf{R})-E(\mathbf{R})\rvert$ and the force-equivariance residual $d_F=\lVert\mathbf{F}(\mathsf{Q}\mathbf{R})-\mathsf{Q}\,\mathbf{F}(\mathbf{R})\rVert$ over the rotations, each normalised by $10^{-3}$ (in eV and eV/{\AA} respectively) before being summed:
\begin{equation}
    \mathcal{S}_\text{Equivariance} = \exp \left( - \left( \frac{\overline{d_E}}{10^{-3} \text{ eV}} + \frac{\overline{d_F}}{10^{-3} \text{ eV/\AA}} \right) \right)
    \label{eq:score_equivariance}
\end{equation}

\subsubsection*{Improper-symmetry stress selection rule}
This probe certifies the improper part of $\mathrm{O}(3)$ through a stress component that the substrate's point group forbids outright.
Under an orthogonal operation $\mathsf{Q}$ the stress transforms as a rank-two tensor, $\sigma(\mathsf{Q}\mathbf{R})=\mathsf{Q}\,\sigma(\mathbf{R})\,\mathsf{Q}^{\!\top}$; for the inversion $\mathsf{P}=-\mathsf{I}$ the two sign factors cancel, giving the exact identity
\begin{equation}
  \sigma(\mathsf{P}\mathbf{R})=\sigma(\mathbf{R}).
  \label{eq:paritystress}
\end{equation}
The substrate is four atoms of a single element on one-fold Wyckoff sites of space group $Pmm2$, one each on $1a\,(0,0,z)$ and $1c\,(\tfrac12,0,z)$ and two on $1d\,(\tfrac12,\tfrac12,z)$ at different heights, with three unequal lattice parameters and generic heights $z_i$, uniformly rescaled so that the nearest-neighbour distance matches the covalent diameter.
In $mm2$ the shear components $\sigma_{xz}$ and $\sigma_{yz}$ are odd under the proper twofold rotation $C_{2z}$ and therefore vanish already for any rotation-equivariant model, whereas $\sigma_{xy}$ is even under $C_{2z}$ and is eliminated only by the mirrors $m_x$ and $m_y$, improper operations with their determinants being $-1$.
The component $\sigma_{xy}$ is thus a pure improper-symmetry residual, while $\sigma_{xz}$ and $\sigma_{yz}$ act as built-in controls.

Operationally, we evaluate the stress of the substrate and of its parity image, obtained by inverting all fractional coordinates through the cell centre (the Bravais lattice itself is centrosymmetric, so the cell is unchanged).
The probe needs two calls in total.
The parity-even average
\begin{equation}
  \bar{\sigma}=\tfrac12\bigl[\sigma(\mathbf{R})+\sigma(\mathsf{P}\mathbf{R})\bigr]
  \label{eq:paritysym}
\end{equation}
restores identity~\eqref{eq:paritystress} by construction and serves as the internal reference: the deviation is the parity-odd remainder of the prediction,
\begin{equation}
  d_\mathrm{P}=\frac{\lVert\sigma(\mathbf{R})-\bar{\sigma}\rVert_F}{\lVert\bar{\sigma}\rVert_F},
  \quad
  \mathcal{S}_\mathrm{P}=\exp\!\bigl(-\,d_\mathrm{P}/\lambda\bigr),
  \quad \lambda=10^{-3},
  \label{eq:score_parity}
\end{equation}
reported together with the component-resolved read-out $\lvert\sigma_{xy}\rvert$ (the improper channel) and $\max(\lvert\sigma_{xz}\rvert,\lvert\sigma_{yz}\rvert)$ (the proper control), each relative to $\lVert\bar{\sigma}\rVert_F$.
For a parity-equivariant model the parity-odd part sits at the floating-point noise floor of the implementation; a parity-blind model leaks into $\sigma_{xy}$ specifically.
Equation~\eqref{eq:paritysym} is simultaneously the practical remedy: symmetrising over the parity image restores the forbidden zeros to machine precision at twice the inference cost.

\subsubsection*{Extensivity and representation invariance}
This probe certifies that the prediction depends on the crystal and not on the bookkeeping used to describe it, across four independent redescriptions.
On the same periodic substrate as the stress probe, we compare the base evaluation against
(i) one mildly sheared unimodular representation, $\mathbf{a}_1'=\mathbf{a}_1+\mathbf{a}_2$, with atoms matched through their wrapped fractional coordinates;
(ii) a $2\times2\times2$ replication, compared per atom;
(iii) a sweep of $16$ rigid translations along a seeded random direction spanning one lattice period, scored by the peak-to-peak energy variation and the largest force change; and
(iv) one random index permutation, with forces compared after undoing the permutation.
Each comparison $c$ yields a per-atom energy residual $\Delta E_c$ (for the supercell after dividing by the replication factor) and a root-mean-square force residual $\Delta F_c$, normalised as in the equivariance probe and summed:
\begin{equation}
  d_\mathrm{rep}=\sum_{c}\left(\frac{\Delta E_c}{10^{-3}\ \text{eV}}
  +\frac{\Delta F_c}{10^{-3}\ \text{eV/\AA}}\right),
  \quad
  \mathcal{S}_\mathrm{rep}=e^{-d_\mathrm{rep}}.
  \label{eq:score_rep}
\end{equation}
The budget is roughly $20$ calls, one of them on the eightfold cell.

\subsubsection*{Hessian null space and acoustic sum rule}
This probe certifies that the Hessian at the model's own stationary point annihilates the rigid translations and rotations, whose form is fixed analytically by geometry.
The null vectors are the rigid translations $\mathbf{t}^{(a)}$ and the infinitesimal rotations $\mathbf{r}^{(a)}_i=\mathbf{e}_a\times(\mathbf{R}_i-\mathbf{R}_{\mathrm{cm}})$,
\begin{equation}
  \mathbf{H}\,\mathbf{v}=\mathbf{0},
  \qquad \mathbf{v}\in\{\mathbf{t}^{(a)},\,\mathbf{r}^{(a)}\}.
  \label{eq:zeromodes}
\end{equation}
For a periodic crystal the same statement at the zone centre is the acoustic sum rule $\sum_j \Phi_{i\alpha,j\beta}=0$.
Operationally, we first relax the cluster (a $13$-atom cluster cut from the element's experimental reference crystal by default) to the model's own minimum, then construct the analytic null vectors directly from the relaxed geometry (three Cartesian translations and, for a non-periodic cluster, the three cross products of Equation~\eqref{eq:zeromodes}, orthonormalised so that $\lVert\mathbf{v}^{(a)}\rVert=1$).
We assemble the dense Hessian from Equation~\eqref{eq:hessfd} at a cost of $6N$ force calls and symmetrise it, $\mathbf{H}\leftarrow\tfrac12(\mathbf{H}+\mathbf{H}^{\!\top})$; all quantities that follow are evaluated on this symmetrised matrix, which is exact for a true Hessian.
We report the continuous residual $\max_a \lVert \mathbf{H}\,\mathbf{v}^{(a)}\rVert/\lVert\mathbf{H}\rVert_F$, which must vanish, together with its un-normalised counterpart $\max_a\lVert\mathbf{H}\,\mathbf{v}^{(a)}\rVert$.
We also count the near-zero eigenvalues of $\mathbf{H}$, taken as those whose magnitude falls below $10^{-2}\lVert\mathbf{H}\rVert_2$, i.e.\ $10^{-2}$ times the largest-magnitude eigenvalue, which must equal $6$ for a generic cluster ($5$ if linear, $3$ acoustic at the zone centre). An eigenvalue that is negative with magnitude exceeding this same threshold at a relaxed minimum flags a spurious soft mode and lowers the score further:
\begin{align}
    d_\text{Mode} &= \bigl\lvert n_0 - n_0^{\mathrm{exp}}\bigr\rvert + n_-, \\
    \mathcal{S}_\text{Null Space}\!&=\!\exp\!\left( \frac{-1}{\lambda}
    \frac{\max_a \lVert \mathbf{H}\,\mathbf{v}^{(a)}\rVert}{\lVert\mathbf{H}\rVert_F} \right)\!\exp(- d_\text{Mode}),
    \label{eq:score_null_space}
\end{align}
where $n_0$ is the number of near-zero eigenvalues, $n_0^{\mathrm{exp}}\in\{6,5,3\}$ the expected count and $n_-$ the number of eigenvalues below $-10^{-2}\lVert\mathbf{H}\rVert_2$.
The default value of the normalisation factor $\lambda$ for this probe is $0.005$.

\subsubsection*{Many-body trimer probe}
This probe combines two independent certifications on one symmetric three-atom complex ($A$--$B$--$C$ ): the decay of the genuine three-body energy, and the symmetry of the force Jacobian in directions transverse to any single bond.
The substrate is built as a symmetric trimer with $A$ and $C$ related by the mirror plane $x=0$ and $B$ on that plane: with an edge length $d$ set from the covalent radius, $A=(-\tfrac{d}{2},0,0)$, $B=(0,h,0)$ with $h=\sqrt{d^2-(d/2)^2}$ and $C=(\tfrac{d}{2},0,0)$.

(i) The many-body-expansion three-body term
\begin{equation}
  \Delta E_3 = E(ABC)-\!\!\sum_{\text{pairs}}\!\! E + \!\!\sum_{\text{atoms}}\!\! E
  \label{eq:threebody}
\end{equation}
is genuine physics at contact and is therefore never scored at contact; instead we recede the apex $B$ from the $AC$ edge along $+y$ over a grid of separations ($16$ points out to $12\,\text{\AA}$ by default).
At each separation we re-evaluate the full trimer and the two pair energies $E(AB)$ and $E(BC)$, while the $AC$ pair and the three monomer energies are computed once, as they do not move.
We then certify that $\Delta E_3(d)\to 0$ monotonically, with the deviation
\begin{align}
  d_\text{TV} &= \left[\frac{\sum_k\lvert\Delta E_3^{(k+1)}-\Delta E_3^{(k)}\rvert}{\lvert\Delta E_3^{(1)}-\Delta E_3^{(N)}\rvert}\!-\!1\!\right]_+, \\
  d_{\Delta E_3} &=\frac{\lvert\Delta E_3^{(N)}\rvert}{\max_k\lvert\Delta E_3^{(k)}\rvert} + d_\text{TV},
  \label{eq:d3}
\end{align}
combining the residual three-body energy at maximal separation, relative to its peak, with a total-variation penalty $[\,\cdot\,]_{+}=\max(0,\cdot)$ that vanishes if and only if the sweep is monotone.

(ii) The transverse term certifies the symmetry of the force Jacobian.
For any configuration $\mathbf{R}\in\mathbb{R}^{3N}$ the model returns a force field $\mathbf{F}(\mathbf{R})$ and its Jacobian collects every first derivative of that field,
\begin{equation}
  \mathsf{J}_{i\alpha,\,j\beta} = \frac{\partial F_{i\alpha}}{\partial R_{j\beta}},
\end{equation}
a $3N\times3N$ matrix whose entry measures how the force on atom $i$ along direction $\alpha$ responds when atom $j$ is displaced along $\beta$. 
Whenever the forces derive from a scalar potential, $\mathbf{F}=-\nabla E$, the Jacobian is exactly minus the Hessian, $\mathsf{J}=-\partial^2 E/\partial\mathbf{R}^2$ and the equality of mixed second derivatives makes it symmetric, $\mathsf{J}=\mathsf{J}^{\!\top}$.
We never form $\mathsf{J}$: the bilinear contraction collapses the test to a scalar,
\begin{align}
  \langle \mathbf{w},\mathsf{J}\mathbf{u}\rangle
  -\langle \mathbf{u},\mathsf{J}\mathbf{w}\rangle
  &= \mathbf{w}^{\!\top}\mathsf{J}\mathbf{u} - \mathbf{u}^{\!\top}\mathsf{J}\mathbf{w} \\
  &= \mathbf{w}^{\!\top}\mathsf{J}\mathbf{u} - \mathbf{w}^{\!\top}\mathsf{J}^{\!\top}\mathbf{u} \\
  &=\langle \mathbf{w},(\mathsf{J}-\mathsf{J}^{\!\top})\mathbf{u}\rangle,
  \label{eq:transverse}
\end{align}
which vanishes for every pair $\mathbf{u},\mathbf{w}$ if and only if $\mathsf{J}$ is symmetric; a small number of random directions thus suffices to expose any asymmetry, with no privileged basis required.
Operationally we draw $n_\perp=8$ seeded pairs of random unit vectors $(\mathbf{u},\mathbf{w})$, evaluate each contraction through the Jacobian--vector product of Equation~\eqref{eq:hvp} (four force calls per pair) and set $d_\perp$ to the mean absolute asymmetry $\lvert\langle\mathbf{w},\mathsf{J}\mathbf{u}\rangle-\langle\mathbf{u},\mathsf{J}\mathbf{w}\rangle\rvert$ in eV/\AA$^2$.
The probe is matrix-free in this sense and its cost is independent of system size beyond the cost of a force call.

The two terms are scored separately and combined as their geometric mean,
\begin{equation}
  \mathcal{S}_\text{Trimer} = \exp\left[-\tfrac12 \left(\frac{d_{\Delta E_3}}{\lambda_{\Delta E_3}}+\frac{d_\perp}{\lambda_\perp}\right)\right],
  \label{eq:score_trimer}
\end{equation}
with a fixed $\lambda_{\Delta E_3}=0.2$ for the three-body term and $\lambda_\perp=0.05$ for the transverse term.

\subsubsection*{Collinear conservativeness}
This probe certifies that the projected force along a bond stretch is minus the energy gradient with respect to that stretch.
Along a one-dimensional bond stretch parameterised by the displacement $s$ of one bonded atom along the bond axis $\hat{\mathbf{u}}$,
\begin{equation}
  \mathbf{F}_j\!\cdot\!\hat{\mathbf{u}} = -\,\frac{\mathrm{d}E}{\mathrm{d}s},
  \label{eq:conservativeness}
\end{equation}
for any conservative model; the projection must be taken on the displaced atom $j$, since $s$ parameterises its coordinate.
We select a bond in the substrate, by default the closest pair in the $13$-atom cluster cut from the element's experimental reference crystal, though any bond may be specified, and displace one of its atoms along the bond axis over a symmetric scan ($\pm0.15\,\text{\AA}$, $31$ points by default).
At each point we record the projected force $\mathbf{F}_j\cdot\hat{\mathbf{u}}$ and the energy and we infer the energy-derived force from a fourth-order central difference,
\begin{equation}
  \left.\frac{\mathrm{d}E}{\mathrm{d}s}\right|_{s} \approx \frac{-E_{s+2h}+8E_{s+h}-8E_{s-h}+E_{s-2h}}{12h}.
  \label{eq:stencil}
\end{equation}
The deviation is the mean absolute mismatch between the two, taken over the interior of the scan,
\begin{equation}
  d_\text{cons}
  = \Big\langle\, \Big| \mathbf{F}_j\!\cdot\!\hat{\mathbf{u}} + \frac{\mathrm{d}E}{\mathrm{d}s} \Big| \,\Big\rangle_{s},
  \quad \mathcal{S}_\text{cons} = \exp\!\Big(\frac{-d_\text{cons}}{\lambda}\Big),
  \label{eq:score_cons}
\end{equation}
which vanishes exactly when Equation~\eqref{eq:conservativeness} holds.
We use $\lambda=0.05$.

\subsubsection*{Maxwell--Betti reciprocity}
This probe certifies the symmetry of the Hessian element by element in the Cartesian basis, as a finite-displacement response experiment.
The reciprocal theorem of linear elasticity states that the response at one degree of freedom to a unit displacement at another equals the reciprocal response; equivalently, the Hessian is symmetric,
\begin{equation}
  \frac{\partial \mathbf{F}_j}{\partial \mathbf{R}_i}
  = \frac{\partial \mathbf{F}_i}{\partial \mathbf{R}_j}.
  \label{eq:betti}
\end{equation}
For a random pair of Cartesian degrees of freedom $(i,j)$, we push along $i$ by $\pm\delta$ and read the induced force at $j$, giving the energy-Hessian element
\begin{equation}
  \mathbf{H}_{ji} = -\frac{\mathbf{F}_j(\mathbf{R}+\delta\mathbf{e}_i)-\mathbf{F}_j(\mathbf{R}-\delta\mathbf{e}_i)}{2\delta},
  \label{eq:betti_fd}
\end{equation}
then push along $j$ and read at $i$ for $\mathbf{H}_{ij}$. The deviation is the mean cross-response asymmetry over a set of $N_p$ such pairs ($12$ by default),
\begin{align}
  d_\text{MB} &= \frac{1}{N_p}\sum_{p=1}^{N_p}\left\lvert \mathbf{H}_{j_p i_p}-\mathbf{H}_{i_p j_p}\right\rvert, \\
  \mathcal{S}_\text{MB} &= \exp\!\left(-\frac{d_\text{MB}}{\lambda}\right),
  \label{eq:score_betti}
\end{align}
with $\lambda=0.02$.

\subsubsection*{Strain--position reciprocity}
This probe certifies the mixed position--strain block of the second-derivative structure, which couples the model's force head to its stress head.
Schwarz's theorem constrains that block on the joint space $(\mathbf{R},\boldsymbol{\varepsilon})$,
\begin{equation}
  V\,\frac{\partial \sigma_{mn}}{\partial R_{j\alpha}}
  =\frac{\partial^{2}E}{\partial R_{j\alpha}\,\partial\varepsilon_{mn}}
  =-\,\frac{\partial F_{j\alpha}}{\partial\varepsilon_{mn}} ,
  \label{eq:crossmaxwell}
\end{equation}
so how the stress responds to an atomic displacement must equal, up to a sign, how the force responds to a strain.
This mixed block is the internal-strain tensor $\Lambda$ of lattice dynamics (the quantity that converts clamped-ion into relaxed-ion elastic constants, $C^{\mathrm{rel}}=C^{\mathrm{clamp}}-\Lambda^{\!\top}\mathbf{H}^{-1}\Lambda$~\cite{internalstrain}).
Operationally, we first relax positions and cell together to the model's own mechanical equilibrium.
On the relaxed periodic substrate we then draw $N_p=6$ random pairs of a Cartesian degree of freedom $j$ and a Voigt component $m$.
The left-hand side of Equation~\eqref{eq:crossmaxwell} follows from displacing $j$ by $\pm\delta$ ($\delta=10^{-3}$~\AA) at fixed cell and central-differencing the predicted stress; the right-hand side from applying the strain $\pm\varepsilon_m$ (with the identical deformation convention as in Equation~\eqref{eq:stress_fd}) and central-differencing the predicted force.
The per-pair residual is relative,
\begin{equation}
  r_p=\frac{\bigl\lvert V\,\partial\sigma_m/\partial R_j+\partial F_j/\partial\varepsilon_m\bigr\rvert}
  {\max\bigl(\lvert V\,\partial\sigma_m/\partial R_j\rvert,\ \lvert\partial F_j/\partial\varepsilon_m\rvert,\ f\bigr)},
  \label{eq:crossresidual}
\end{equation}
with a small floor $f$ guarding vanishing responses and $d_\Lambda=\langle r_p\rangle_p$.

At marginal extra cost the strain--strain block completes the picture: assembling $B_{mn}=\partial\sigma_m/\partial\varepsilon_n$ from central differences of the predicted stress under the six Voigt strains ($12$ calls) tests the stress head's own integrability through its Voigt symmetry, $d_B=\lVert B-B^{\!\top}\rVert_F/\lVert B\rVert_F$, independently of the energy.
The two deviations are combined as their geometric mean,
\begin{equation}
  \mathcal{S}_\Lambda=\exp\!\left[-\tfrac12\left(\frac{d_\Lambda}{\lambda}+\frac{d_B}{\lambda}\right)\right],
  \quad \lambda=0.05 ,
  \label{eq:score_crossmaxwell}
\end{equation}
at a cost of $4N_p+12=36$ calls beyond the preceding relaxation.

\subsubsection*{Stress--gradient consistency}
This probe certifies that the model's stress head is the volume-normalised strain derivative of its own energy,
\begin{equation}
  \sigma_{ij}=\frac{1}{V}\,\frac{\partial E}{\partial \varepsilon_{ij}} .
  \label{eq:stress}
\end{equation}
The substrate is a small periodic crystal of the element (an \textsc{ASE} reference structure where one exists, otherwise a covalent-radius-sized fcc cell).
We read the model's own stress once, then for each of the six independent Voigt strains $\varepsilon_k$ apply the deformation gradient $\mathsf{F}=\mathsf{I}+\boldsymbol{\varepsilon}_k$ to the cell with the atoms scaled affinely, evaluate the energy at $\pm\delta$ and form the central-difference stress component
\begin{equation}
  \sigma^{\mathrm{FD}}_{k}=\frac{E(+\varepsilon_k)-E(-\varepsilon_k)}{2\delta\,V}.
  \label{eq:stress_fd}
\end{equation}
The shear components enter the symmetric strain tensor at half magnitude, so the predicted and finite-difference vectors are directly comparable.
The deviation is the relative Frobenius residual of the full tensor,
\begin{equation}
  d_\sigma = \frac{\lVert\boldsymbol{\sigma}^{\mathrm{pred}}-\boldsymbol{\sigma}^{\mathrm{FD}}\rVert_F}{\lVert\boldsymbol{\sigma}^{\mathrm{FD}}\rVert_F},
  \quad
  \mathcal{S}_\sigma = \exp\bigl(-\,d_\sigma/\lambda\bigr),
  \label{eq:score_stress}
\end{equation}
with $\lambda=0.05$.

\subsubsection*{Time-reversibility}
This probe certifies that a microcanonical trajectory retraces itself when the momenta are reversed, which a smooth conservative surface propagated with a symplectic integrator must do.
\begin{align}
  \mathbf{R}_0 &\xrightarrow{N} \mathbf{R}_1
  \xrightarrow{\;-\mathbf{v},\,N\;} \mathbf{R}_0', \\
  d_\mathrm{rt} &=\sqrt{\tfrac{1}{N_\mathrm{at}}\textstyle\sum_i \lVert \mathbf{R}_{0,i}'-\mathbf{R}_{0,i}\rVert^{2}} \to 0 .
  \label{eq:reversibility}
\end{align}
We initialise seeded Maxwell--Boltzmann velocities on the cluster (the $13$-atom cluster cut from the element's experimental reference crystal and $30$~K by default), propagate $N$ velocity-Verlet steps (here $N=100$ at $1$~fs) in the microcanonical ensemble, reverse all momenta, propagate $N$ steps back and report the per-atom root-mean-square return displacement $d_\mathrm{rt}$, which is the scored quantity:
\begin{equation}
  \mathcal{S}_\mathrm{rev}=\exp\!\big(-\,d_\mathrm{rt}/\lambda\big),
  \qquad \lambda=0.001.
  \label{eq:score_reversibility}
\end{equation}
The secular energy drift is recorded separately as the slope of the total energy along the trajectory.

\subsubsection*{Configurational temperature}
This probe certifies that the temperature read from the spatial fluctuations of the force agrees with the kinetic temperature, which holds only if the trajectory samples the Boltzmann distribution.
A purely configurational thermometer relates those fluctuations to the mean curvature~\cite{rugh1997,butler1998},
\begin{equation}
  k_{B}\,T_{\mathrm{conf}}
  = \frac{\langle \lVert\nabla U\rVert^{2}\rangle}{\langle \nabla^{2}U\rangle}
  = \frac{\langle \lVert\mathbf{F}\rVert^{2}\rangle}{\langle \operatorname{Tr}\mathbf{H}\rangle},
  \qquad \frac{T_{\mathrm{conf}}}{T_{\mathrm{kin}}}\to 1 ,
  \label{eq:configtemp}
\end{equation}
the two coinciding only when the trajectory samples the Boltzmann distribution.
Operationally we relax to the model's own minimum, run a Langevin trajectory ($13$-atom cluster cut from the element's experimental reference crystal at $30$~K and $1{,}100$ sampling steps at $1.0$~fs by default), equilibrate ($400$~steps) and then average both quantities over every 10th snapshot by default: the trace is estimated matrix-free by a Hutchinson average~\cite{hutchinson1989}: $\operatorname{Tr}\mathbf{H}\!\approx\!\langle \mathbf{z}^{\!\top}\mathbf{H}\mathbf{z}\rangle$ over a few seeded $\pm1$ vectors (four vectors by default).
The kinetic temperature is taken over the full $3N$ degrees of freedom: $T_\mathrm{kin} = 2\langle E_\mathrm{kin} \rangle / (3Nk_B)$.
The deviation and its score are
\begin{equation}
  d_T = \left\lvert\frac{T_{\mathrm{conf}}}{T_{\mathrm{kin}}}-1\right\rvert,
  \quad
  \mathcal{S}_\text{conf} = \exp\bigl(-\,d_T\bigr).
  \label{eq:score_configtemp}
\end{equation}

\subsubsection*{Virial theorem}
This probe certifies the Clausius virial balance on the same trajectory, using positions and forces only and no curvature at all.
The theorem is exact for any bound classical system in a stationary state, harmonic or not~\cite{clausius1870}: the virial of the internal forces balances twice the internal kinetic energy,
\begin{equation}
  \rho_{\mathrm{vir}} = \frac{-\big\langle \sum_i (\mathbf{R}_i - \mathbf{R}_i^{\mathrm{eq}})\cdot\mathbf{F}_i \big\rangle}{(3N-6)\,k_B T^{\mathrm{int}}_{\mathrm{kin}}} \;\to\; 1,
  \label{eq:virial}
\end{equation}
where $\mathbf{R}^{\mathrm{eq}}$ is the relaxed minimum from which the trajectory starts.
The probe is evaluated on the Langevin trajectory produced for the configurational temperature probe: by default at every fourth snapshot the numerator is formed from the forces and the displacements $\mathbf{R}_i-\mathbf{R}_i^{\mathrm{eq}}$ from the relaxed minimum. 
The internal kinetic temperature is obtained after projecting out the rigid translation and rotation of the cluster (subtracting the centre-of-mass velocity and the rigid rotation $\boldsymbol{\omega}\times(\mathbf{R}_i-\mathbf{R}_\mathrm{cm})$, with $\boldsymbol{\omega}$ from the instantaneous inertia tensor and angular momentum), since the six unconfined degrees of freedom contribute kinetic energy but no confining virial.
The deviation and score follow the configurational-temperature convention,
\begin{equation}
  d_\mathrm{vir}=\lvert\rho_\mathrm{vir}-1\rvert,
  \quad
  \mathcal{S}_\mathrm{vir}=\exp(-\,d_\mathrm{vir}).
  \label{eq:score_virial}
\end{equation}

\subsubsection*{Per-mode equipartition}
This probe certifies, mode by mode, that the static curvature of the surface and the amplitude with which the dynamics populates each mode are consistent with equipartition.
Near a minimum the energy surface is, to leading order, a sum of independent harmonic oscillators in the mass-weighted normal coordinates $Q_\alpha$ and the classical equipartition theorem fixes the thermal amplitude of every non-zero mode:
\begin{equation}
  R_\alpha = \frac{\omega_\alpha^{2}\,\langle Q_\alpha^{2}\rangle}{k_{B}T}\to 1 .
  \label{eq:equipartition}
\end{equation}
The curvature channel assembles the finite-difference Hessian ($6N$ force calls), mass-weights it to the dynamical matrix $\mathbf{D}=\mathbf{M}^{-0.5}\mathbf{H}\,\mathbf{M}^{-0.5}$ and diagonalises it: the eigenvalues give $\{\omega_\alpha^{2}\}$ and the eigenvectors $\{\mathbf{e}_\alpha\}$ the projection basis, retaining only stable modes: positive eigenvalues above a small relative tolerance, which discards the six zero modes and any spurious imaginary ones.
For every configuration in the Langevin trajectory produced for the configurational temperature probe $\langle Q_\alpha^{2}\rangle$ is accumulated from the mass-weighted displacements projected onto each mode, $Q_\alpha=\mathbf{e}_\alpha^{\!\top}\mathbf{M}^{0.5}(\mathbf{R}-\mathbf{R}^{\mathrm{eq}})$.
Each per-mode ratio carries an irreducible finite-sampling scatter even for a perfect model, so we score not the spread but how well the spectrum centres on unity, using the sampling-robust median:
\begin{align}
  d_\mathrm{eq} &= \bigl\lvert \operatorname{median}_\alpha (R_\alpha) - 1 \bigr\rvert,
  \\
  \mathcal{S}_\mathrm{eq} &= \exp\left(-\frac{d_\mathrm{eq}}{\lambda}\right),
  \quad \lambda = 1.0 .
  \label{eq:score_equipartition}
\end{align}

\subsubsection*{Bond-deformation smoothness}
This probe certifies that a bond scan from compression through dissociation traces a smooth single well, free of the localised artefacts that no aggregate error metric reveals.
We scan the closest atom pair (in the $13$-atom cluster cut from the element's experimental reference crystal) from moderate compression out into the dissociation tail, an absolute stretch beyond equilibrium, recording the projected force on the displaced atom and the energy at each of $M$ points ($81$~steps by default).
Writing $F_k$ for the projected force and $\delta_k=F_{k+1}-F_k$ for its increments along the scan, we form four diagnostics, each a penalty that vanishes for an ideal well:
(i) a force spike $\rho$, the largest local deviation of an increment from the mean of its neighbours, $g_k=\delta_k-\tfrac12(\delta_{k-1}+\delta_{k+1})$, normalised by the typical step, so that a smooth force gives an $\mathcal{O}(1)$ ratio while a step, kink, or isolated outlier produces a large value;
(ii) the number of force sign reversals in excess of the single crossing;
(iii) the number of energy minima in excess of one;
and (iv) the force tortuosity, the total variation of the projected force normalised by its range, zero for a monotone force.
These combine into a single defect:
\begin{align}
  \rho&=\frac{\max_k\lvert g_k\rvert}{\operatorname{median}_k\lvert\delta_k\rvert}, \quad
  \mathrm{TV}(F)=\sum_k\lvert\delta_k\rvert, \\
  \Delta F&=\max_k F_k-\min_k F_k, \\
  d_\mathrm{spike}&=w\,[\rho-\rho_0]_+,\quad
  d_\mathrm{flip}=[n_\mathrm{flip}-1]_+,\\
  d_\mathrm{min}&=[n_\mathrm{min}-1]_+,\quad
  d_\mathrm{TV}=\Big[\tfrac{\mathrm{TV}(F)}{\Delta F}-1\Big]_+, \\
  \mathcal{S}_\mathrm{smooth}
  &=\exp\!\left(-\,\frac{d_\mathrm{spike}+d_\mathrm{flip}+d_\mathrm{min}+d_\mathrm{TV}}{\lambda}\right),
  \label{eq:score_smoothness}
\end{align}
with $[\,\cdot\,]_+=\max(0,\cdot)$, the spike ratio $\rho$ entering only above a baseline $\rho_0=4$ and at reduced weight $w=\tfrac14$, the other three penalties at unit weight and $\lambda=1$.

\subsection*{Aggregation}
Each probe returns the same result object: the exact target, the non-negative deviation $d$, the normalised score $\mathcal{S}=e^{-d/\lambda}$ with its recorded scale $\lambda$ and the number of model calls consumed.
The probe scores are combined into an overall certification by their geometric mean over all $\mathcal{P}$ applicable probes,
\begin{equation}
  \mathcal{S}_\mathrm{overall}=\Bigl(\prod_{p=1}^{\mathcal{P}}\mathcal{S}_p\Bigr)^{1/\mathcal{P}},
  \label{eq:overall}
\end{equation}
which, unlike an arithmetic mean, cannot hide a single failing axis behind many passed gates.
The probe scores $\mathcal{S}_p$ are clipped to $[10^{-12},1]$ before the mean is taken.
Probes that do not apply to a model (for example, the stress-based probes for a model without a stress head) are skipped and excluded from $\mathcal{P}$.
For the benchmark of Figure~\ref{fig:heatmap}, every probe is run once per element on that element's substrates, and the ten per-element scores of a probe are combined into the single entry shown in the figure by their arithmetic mean; the overall certification of Equation~(\ref{eq:overall}) is then taken over these per-probe entries.

\section*{Data Availability}

The exact version of the certification suite and the corresponding probe results for the investigated UMLIPs in this study and the fine-tuning study are available at \url{10.5281/zenodo.22731106}.
The detailed results from the UMLIP investigations are additionally presented in an interactive way at \url{https://jhaens.github.io/pescert-bench}.

\section*{Code Availability}

The reference-free certification suite \textsc{pescert} is available at \url{github.com/jhaens/pescert}.
The used third-party potentials 
\textsc{Allegro}~\cite{allegro,nequi_allegro_software},
\textsc{AllScaIP}~\cite{allattention,omol},
\textsc{CHGNet}~\cite{chgnet},
\textsc{DeepMD}~\cite{dpa, dpa3, dpa4, dpa-4c},
\textsc{EquFlash}~\cite{equflash},
\textsc{Equiformer}~\cite{equiformer,equiformerv2,equiformerv3,dens},
\textsc{eSEN}~\cite{esen, omol},
\textsc{GRACE}~\cite{grace_1,grace_2},
\textsc{M3GNet}~\cite{m3gnet},
\textsc{MACE}~\cite{mace,macemp0,mace_2,mace_omat_mh},
\textsc{MatRIS}~\cite{matris},
\textsc{MatterSim}~\cite{mattersim}, 
\textsc{NEP89}~\cite{nep89},
\textsc{Nequip}~\cite{batzner2022,nequi_allegro_software},
\textsc{ORB}~\cite{orb,orbv3},
\textsc{PET}~\cite{pet_mad,pet_oam,metatrain},
\textsc{Prophet}~\cite{elementa_dataset}, % model not published until now
\textsc{SevenNet}~\cite{sevennet,sevennet_mf,sevennet_omni,sevennet-nano},
\textsc{TACE}~\cite{tace}, 
\textsc{TECE}~\cite{tece},
and \textsc{UMA}~\cite{uma,omol}
are available from their respective repositories.

\section*{Acknowledgments}

We thank the staff of the Compute Center of the Technische Universit{\"a}t Ilmenau, especially Mr.~Henning~Schwanbeck for providing an excellent research environment.
J.~H{\"a}nseroth thanks E.~Pabst, D.~Sebastiani and J.~L.~Wolf for their valuable feedback on the first draft of the manuscript.
J.~H{\"a}nseroth thanks M.~Gro{\ss}mann and E.~Pabst for the discussions concerning the improper symmetry probe.
This work is supported by the doctoral scholarship of the German Academic Scholarship foundation, the Carl-Zeiss-Stiftung (SustEnMat, funding code: P2023-02-008), the Thüringer Aufbaubank (TAB) (KapMemLyse, grant no.~2024 FGR 0081 / 0082) and the European Social Fund Plus (ESF+).

J.~H{\"a}nseroth thanks S.~M.~Blau and B.~M.~Wood for the insightful discussions on the importance of reliable evaluations in the MLIP age that were a huge motivation for this work.

\section*{Competing interests}

The authors declare no competing interests.

\section*{Author contributions}
J.H.~conceived the study, implemented the certification suite and performed all calculations.
J.H.~analysed the data.
J.H.~produced all figures and wrote the first draft of the manuscript.
C.D.~supervised the work.
Both authors revised and approved the manuscript.

\bibliography{bibliography.bib}

\end{document}

\author{Jonas H{\"a}nseroth}
\email{jonas.haenseroth@tu-ilmenau.de}
\affiliation{Theoretical Solid State Physics, Institute of Physics, Technische Universität Ilmenau, 98693 Ilmenau, Germany}

\author{Christian Dre{\ss}ler}
\affiliation{Theoretical Solid State Physics, Institute of Physics, Technische Universität Ilmenau, 98693 Ilmenau, Germany}

\title{Supplementary Information for ``Reference-free certification of machine-learning interatomic potentials''}
\date{\today}

\maketitle

All abbreviations are defined in the main text; abbreviations used only here are defined at first use.
Throughout, ``MAE'' denotes mean absolute error (energies per atom, in eV; forces in eV/{\AA}), ``DOF'' a Cartesian degree of freedom, ``COM'' the centre of mass, ``NVE'' the microcanonical ensemble, ``LJ'' the Lennard-Jones potential, and ``PES'' the potential-energy surface.
Units follow the atomic-simulation environment (\textsc{ASE}): energy in eV, length in {\AA}, force in eV/{\AA}, stress in eV/{\AA}$^3$, temperature in K.

\tableofcontents
\newpage
\clearpage

\newpage
\section*{Supplementary Note 1: The analytic Lennard-Jones testbed}

Each of the fourteen probes of the main text is a small, self-contained routine that interacts with a potential only through its energy, forces, and (where available) stress.
This document validates every probe on a controlled analytic testbed for which the correct answer is known in closed form, and, for each probe, restates exactly what the implementation computes.

{\begin{center} \textbf{Toy testbed} \end{center}}

The central difficulty in validating a probe is that any real machine-learning potential is itself imperfect, so a nonzero deviation cannot be cleanly attributed to the probe or to the model.
We therefore validate against the original Lennard-Jones potential,
\begin{equation}
  U_{\mathrm{LJ}}(r) = 4\varepsilon\!\left[\left(\frac{\sigma}{r}\right)^{12}-\left(\frac{\sigma}{r}\right)^{6}\right],
  \qquad \varepsilon=\sigma=1,
  \label{eq:lj}
\end{equation}
with the cutoff $r_c=8\,\sigma$ placed far beyond any relevant interaction, so that the reference is smooth, exactly conservative, exactly rotation-, translation-, inversion-, and permutation-invariant, exactly pairwise, and exactly reproduces its own stress.
It is, in other words, an ideal model, and every probe must return a near-perfect score on it.
This is the first half of every validation below: the reference row confirms that the probe does not penalise a correct potential (its residual sits at the level of finite-difference or finite-sampling noise).

The second, more informative half uses modulated Lennard-Jones potentials: the reference with one specific, physically interpretable defect added on top.
Each modulation is designed as an adversary for a single probe.
For several probes we additionally include a harmless modulation with a worse energy or force MAE that respects the identity and passes.
Together these realise the litmus test of the main text: a model can obtain an excellent (here sometimes exactly zero) test error and still fail the check, and a larger error does not by itself trip the check.
Because every modulation is a thin wrapper around the same LJ base, its energy and force MAE against the reference is directly quotable, and we quote it in every demonstration (measured over randomly displaced copies of the respective substrate).

{\begin{center} \textbf{Substrates} \end{center}}

The demonstrations use the four chemistry-independent substrates of the main-text suite, populated with argon so the LJ potential applies:
\begin{itemize}
    \item[(i)] a $13$-atom cluster cut from the element's experimental reference crystal (used as substrate for the equivariance, null-space, Maxwell--Betti, reversibility, and all statistical-mechanical probes);
    \item[(ii)] a symmetric mirror-related trimer at the same scaling (used as substrate for the many-body probe);
    \item[(iii)] a periodic fcc cell at the LJ nearest-neighbour distance (used as substrate for the stress--gradient probe); and
    \item[(iv)] the four-atom low-symmetry $Pmm2$ crystal of the main text (used as substrate for the improper-symmetry, representation, and strain--position probes).
\end{itemize}

The two one-dimensional bond probes (collinear conservativeness, smoothness) use an analytic two-atom dimer whose radial energy $e(r)$ and radial slope $g(r)=\mathrm{d}E/\mathrm{d}r$ can be prescribed independently.

{\begin{center} \textbf{Single probe reports} \end{center}}

Every probe returns the same result object: the exact target, a non-negative deviation $d$ from it, the normalised score $\mathcal{S}=\exp(-d/\lambda)$ with its recorded scale $\lambda$, the number of model calls consumed, an optional Boolean gate flag (\texttt{PASS}/\texttt{FAIL} when $d$ sits below/above a few times $\lambda$), and a dictionary of diagnostics.
A deviation equal to $\lambda$ maps to $\mathcal{S}=1/e\approx0.37$, so $\lambda$ is the interpretable sensitivity of each probe.
In every demonstration below we quote the score, the raw deviation, and the key diagnostics exactly as the routine reports them; probe settings differ from the library defaults only where stated (denser scan grids for plotting, and a rescaled temperature for the reversibility probe, since $k_BT$ must be compared with $\varepsilon=1$~eV here rather than with a real material's well depth).

{\begin{center} \textbf{Finite-difference errors} \end{center}}

Because the Lennard-Jones control is analytic and exactly conservative, every residual a probe reports for it is finite-difference error and nothing else: truncation of the stencil, plus round-off in the energies and forces the potential returns.
Running the suite's own stencils on it therefore measures the floor beneath every reported defect.

In Table~\ref{tab:si_fd_floor} the finite-difference errors are shown for the $13$-atom silicon cluster at the two working precisions the suite encounters.
The step for \texttt{float32} itself is coarser, because the model's noise is some nine orders larger; and the forces entering the difference carry only \texttt{float32} significance.
For the collinear conservativeness probe the scan range is a physical choice, so the grid is thinned rather than shrunk: $31$ points at $0.0100$~{\AA} (\texttt{float64}) become $9$ points at $0.0375$~{\AA} (\texttt{float32}).
The Hessian--vector products use $\epsilon=0.0010$~{\AA} at \texttt{float64} and $\epsilon=0.0106$~{\AA} at \texttt{float32}.

\begin{table}[h]
    \centering
    \caption{\textbf{Finite-difference floor on the analytic control.}
    Errors of the suite's own stencils on the $13$-atom Lennard-Jones silicon cluster.
    The conservativeness entry is the mean residual of the fourth-order central difference of the energy against the analytic force over the scan; the Hessian--vector entries are medians over $16$ random unit directions, measured against a Richardson-extrapolated reference.}
    \label{tab:si_fd_floor}
    \begin{tabular}{l c c}
        \toprule
        Quantity & \texttt{float64} & \texttt{float32} \\
        \midrule
        Conservativeness, 4th-order central (eV/{\AA})   & $3.3\times10^{-6}$ & $6.0\times10^{-4}$ \\
        \quad relative to the force magnitude            & $1.9\times10^{-6}$ & $4.1\times10^{-4}$ \\
        Hessian--vector product, absolute (eV/{\AA}$^2$) & $1.3\times10^{-5}$ & $1.4\times10^{-3}$ \\
        \quad relative                                   & $1.6\times10^{-6}$ & $1.8\times10^{-4}$ \\
        \bottomrule
    \end{tabular}
\end{table}

\clearpage

\section*{Supplementary Note 2: Rotational equivariance}

\textbf{Identity and target.}
For any rotation $\mathsf{Q}\in\mathrm{SO}(3)$ the energy is invariant and the forces rotate with the structure, $E(\mathsf{Q}\mathbf{R})=E(\mathbf{R})$ and $\mathbf{F}(\mathsf{Q}\mathbf{R})=\mathsf{Q}\,\mathbf{F}(\mathbf{R})$; both residuals have target $0$.

\textbf{What the implementation computes.}
On the 13-atom cluster the routine evaluates $E(\mathbf{R})$ and $\mathbf{F}(\mathbf{R})$ once, draws $n_{\mathrm{rot}}=8$ proper rotations from a seeded random stream (uniform axis and angle, via Rodrigues' formula), rotates the cluster rigidly about its COM, and re-evaluates.
The energy residual $\overline{d_E}=\langle|E(\mathsf{Q}\mathbf{R})-E(\mathbf{R})|\rangle$ and the force residual $\overline{d_F}=\langle\lVert\mathbf{F}(\mathsf{Q}\mathbf{R})-\mathsf{Q}\,\mathbf{F}(\mathbf{R})\rVert\rangle$ are each normalised by $10^{-3}$ (eV, eV/{\AA}) and summed into $d$, scored with $\lambda=1$.
Cost: $1+n_{\mathrm{rot}}=9$ calls; the gate passes when both residuals sit below $5\times10^{-3}$.

\begin{figure}[htb]
    \centering
    \includegraphics{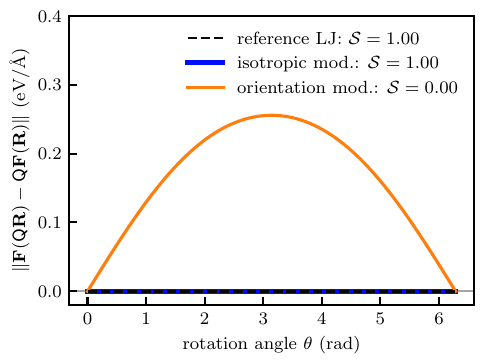}
    \caption{\textbf{Rotational equivariance.} 
    Force-equivariance residual $\lVert\mathbf{F}(\mathsf{Q}\mathbf{R})-\mathsf{Q}\,\mathbf{F}(\mathbf{R})\rVert$ as the 13-atom cluster is rotated about a fixed axis through its COM. 
    The harmless isotropic modulation (blue), though less accurate by energy MAE, keeps the residual at machine precision and scores $1.00$; the deceptive orientation-breaking modulation (orange), though more accurate, produces a large residual and scores $0.00$. 
    Its energy residual is zero by the isotropy of the cluster, so only the force channel exposes it.}
    \label{fig:si_equivariance}
\end{figure}

\textbf{Toy demonstration.} 
The deceptive candidate adds a conservative absolute-orientation term $c\sum_i(\mathbf{R}_i\cdot\hat{\mathbf{n}})^2$ ($c=0.06$, $\hat{\mathbf{n}}\parallel(1,1,1)$), which references a fixed laboratory axis.
On the cluster the added energy is constant under rotations about the COM, so the energy residual is numerically zero ($8.9\times10^{-16}$); the forces, however, do not rotate with the structure ($\overline{d_F}=0.21$).
The harmless candidate adds $\tfrac12 c'\sum_i\lVert\mathbf{R}_i-\mathbf{R}_{\mathrm{cm}}\rVert^2$ ($c'=0.15$), built from distances to the COM and hence genuinely invariant.
The probe reports (Supplementary Figure~\ref{fig:si_equivariance}): reference LJ $\mathcal{S}=1.000$ ($d=3.3\times10^{-11}$); deceptive orientation modulation $\mathcal{S}=0.00$ ($d=208$, gate \texttt{FAIL}) at an energy MAE of only $0.024$; harmless isotropic modulation $\mathcal{S}=1.000$ at a nearly four times larger energy MAE of $0.089$.

\clearpage
\section*{Supplementary Note 3: Improper-symmetry stress selection rule}

\textbf{Identity and target.} 
The exact surface is invariant under the full orthogonal group, so under the inversion $\mathsf{P}=-\mathsf{I}$ the stress satisfies $\sigma(\mathsf{P}\mathbf{R})=\sigma(\mathbf{R})$ exactly; the parity-odd remainder has target $0$.

\textbf{What the implementation computes.} 
On the four-atom $Pmm2$ crystal the routine evaluates the stress of the substrate and of its parity image (all fractional coordinates inverted through the cell centre; the Bravais lattice is centrosymmetric, so the cell is unchanged); two calls in total.
The parity-even average $\bar{\sigma}=\tfrac12[\sigma(\mathbf{R})+\sigma(\mathsf{P}\mathbf{R})]$ restores the identity by construction; the deviation is $d_{\mathrm{P}}=\lVert\sigma(\mathbf{R})-\bar{\sigma}\rVert_F/\lVert\bar{\sigma}\rVert_F$, scored with $\lambda=10^{-3}$.
It reports the component-resolved read-out alongside: in point group $mm2$ the shear $\sigma_{xy}$ is even under the proper $C_{2z}$ and is forbidden only by the improper mirrors (the pure improper channel), while $\sigma_{xz},\sigma_{yz}$ vanish for any rotation-equivariant model (built-in proper controls).

\begin{figure}[htb]
    \centering
    \includegraphics{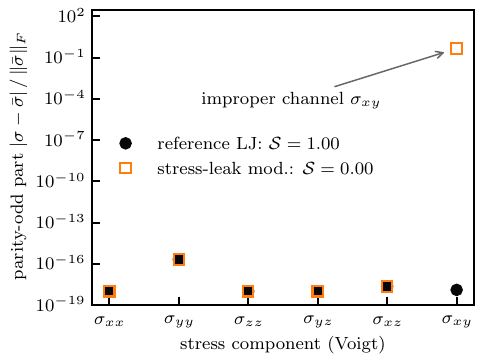}
    \caption{\textbf{Improper-symmetry stress selection rule.} 
    Parity-odd remainder of each Voigt stress component, relative to $\lVert\bar{\sigma}\rVert_F$, on the $Pmm2$ crystal. 
    The reference LJ (black) sits at the floating-point floor in every component. The stress-leak modulation (orange), whose energy and force errors are exactly zero, leaks only into the improper channel $\sigma_{xy}$, while the proper controls $\sigma_{xz},\sigma_{yz}$ remain clean.}
    \label{fig:si_parity}
\end{figure}

\textbf{Toy demonstration.} 
The deceptive candidate leaks a parity-odd term into the improper channel: it adds $a\sum_i\sin(2\pi z_i)$ (fractional $z_i$, $a=0.02$) to the reported $\sigma_{xy}$ only.
This functional flips sign under inversion of the fractional coordinates, so the leaked part obeys $\sigma_{xy}(\mathsf{P}\mathbf{R})=-\sigma_{xy}(\mathbf{R})$; energies and forces are untouched, so the candidate's energy and force MAE are exactly zero.
The probe reports (Supplementary Figure~\ref{fig:si_parity}): reference LJ $\mathcal{S}=1.000$ ($d_{\mathrm{P}}=2.1\times10^{-16}$); stress-leak modulation $\mathcal{S}=0.00$ ($d_{\mathrm{P}}=0.64$, gate \texttt{FAIL}).
The channel resolution localises the failure: the improper channel carries $|\sigma_{xy}-\bar{\sigma}_{xy}|/\lVert\bar{\sigma}\rVert_F=0.45$ while the proper controls stay at $2.3\times10^{-18}$, so the defect is attributed to the improper operations specifically, exactly as designed.

\clearpage
\section*{Supplementary Note 4: Extensivity and representation invariance}

\textbf{Identity and target.} 
Energy, forces, and stress must be invariant under a unimodular change of the lattice basis, extensive under supercell replication, invariant under rigid translation of all atoms modulo the cell, and invariant under permutation of atom indices.
The summed normalised residual $d_{\mathrm{rep}}$ has target $0$.

\textbf{What the implementation computes.} 
On the $Pmm2$ crystal the routine compares the base evaluation against (i) the sheared unimodular representation $\mathbf{a}_1'=\mathbf{a}_1+\mathbf{a}_2$ (atoms matched through wrapped fractional coordinates), (ii) a $2\times2\times2$ replication compared per atom, (iii) a sweep of $16$ rigid translations along a seeded random direction spanning one lattice period (scored by the peak-to-peak energy and the largest force change), and (iv) one random index permutation, undone before comparing forces.
Each comparison yields a per-atom energy residual and an RMS force residual, normalised by $10^{-3}$ (eV, eV/{\AA}) and summed into $d_{\mathrm{rep}}$, with $\mathcal{S}_{\mathrm{rep}}=e^{-d_{\mathrm{rep}}}$; the budget is $20$~calls.

\begin{figure}[htb]
    \centering
    \includegraphics{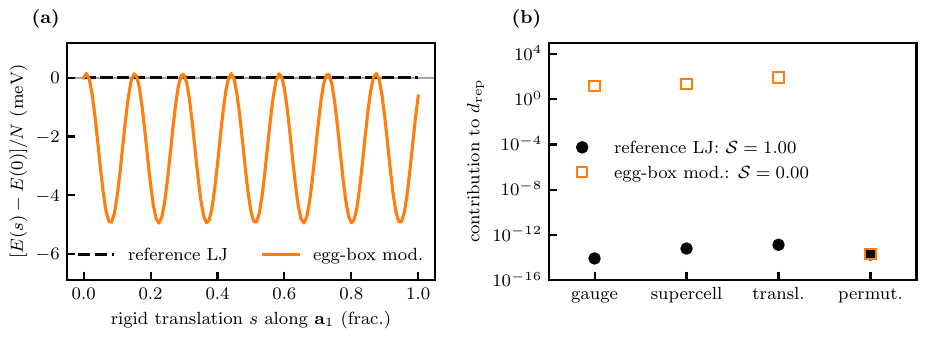}
    \caption{\textbf{Extensivity and representation invariance.} (a)~Per-atom energy along a rigid translation of all atoms by one lattice period: exactly flat for the reference LJ (black), oscillating for the egg-box modulation (orange). 
    (b)~Contribution of each comparison to $d_{\mathrm{rep}}$ (log scale). 
    The egg-box modulation trips the gauge, supercell, and translation channels but leaves the permutation channel at machine precision, which it respects by construction.}
    \label{fig:si_representation}
\end{figure}

\textbf{Toy demonstration.}
The deceptive candidate imports the egg-box artefact of grid-based codes as an analytic modulation: it adds $A\sum_{i}\sum_{\alpha}[1-\cos(2\pi R_{i\alpha}/h)]$ with amplitude $A=5$~meV and grid spacing $h=0.53$~{\AA} incommensurate with the lattice, a smooth, conservative term (forces are its exact gradient) that pins the energy to an absolute Cartesian grid.
Its accuracy cost is small (energy MAE $0.015$ per atom, force MAE $0.032$), yet it breaks three of the four invariances at once.
The probe reports (Supplementary Figure~\ref{fig:si_representation}): reference LJ $\mathcal{S}=1.000$ ($d_{\mathrm{rep}}=2.2\times10^{-13}$); egg-box modulation $\mathcal{S}=0.00$ ($d_{\mathrm{rep}}=122$), with the contributions resolving which invariances fail: unimodular gauge $14.3$, supercell $22.1$, translation $85.7$, permutation $2.1\times10^{-14}$.
The permutation channel, which the modulation respects, stays at machine precision and acts as the built-in control, confirming that the probe attributes the defect and does not merely accumulate noise.

\clearpage
\section*{Supplementary Note 5: Hessian null space and acoustic sum rule}

\textbf{Identity and target.} 
At a stationary point the Hessian must annihilate the rigid-body modes: three translations and, for a finite cluster, three infinitesimal rotations $\mathbf{r}^{(a)}_i=\mathbf{e}_a\times(\mathbf{R}_i-\mathbf{R}_{\mathrm{cm}})$.
The continuous residual $\max_a\lVert\mathbf{H}\mathbf{v}^{(a)}\rVert/\lVert\mathbf{H}\rVert_F$ has target $0$, and the near-zero eigenvalue count must equal $6$ (generic cluster; $5$ linear, $3$ crystal at $\Gamma$).

\textbf{What the implementation computes.} 
The routine relaxes the cluster with the model's own forces (BFGS, $F_{\max}=10^{-3}$~eV/{\AA}), builds the analytic null vectors from the relaxed geometry, assembles the dense Hessian matrix-free from central differences of forces ($6N$ calls, Equation~(1) of the main text), and symmetrises it.
It reports the continuous residual, the count $n_0$ of eigenvalues with $|\lambda_i|<10^{-2}\lVert\mathbf{H}\rVert_2$, and the count $n_-$ of negative eigenvalues beyond the same threshold; the score multiplies the continuous term ($\lambda=5\times10^{-3}$) by $\exp[-(|n_0-n_0^{\mathrm{exp}}|+n_-)]$.

\begin{figure}[htb]
    \centering
    \includegraphics{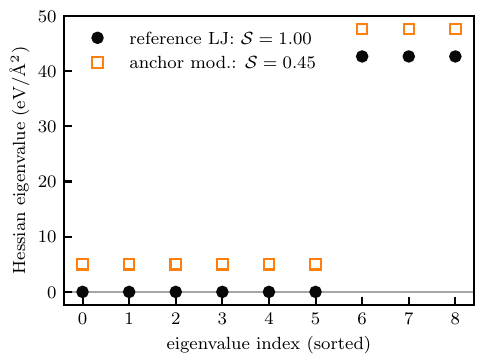}
    \caption{\textbf{Hessian null space.} 
    Lowest sorted Hessian eigenvalues at the relaxed minimum. The reference LJ (black) has six eigenvalues at zero (the exact rigid-body null space). 
    The anchor modulation (orange) shifts every eigenvalue up by the spring constant $c=5$~eV/{\AA}$^2$: the near-zero count is unchanged, but the continuous residual $\max_a\lVert\mathbf{H}\mathbf{v}^{(a)}\rVert/\lVert\mathbf{H}\rVert_F$ is finite and the score drops to $0.45$.}
    \label{fig:si_zeromodes}
\end{figure}

\textbf{Toy demonstration.} 
The deceptive candidate adds a weak absolute-position spring $\tfrac12 c\sum_i\lVert\mathbf{R}_i-\mathbf{a}_i\rVert^2$ ($c=5$~eV/{\AA}$^2$) anchored at the relaxed geometry $\mathbf{a}_i$.
At the minimum the spring exerts no force, so geometry and energies nearby are barely changed (energy MAE $0.012$, force MAE $0.16$); but it adds the constant $c$ to every Hessian eigenvalue, lifting the six rigid-body modes that must be exactly flat (Supplementary Figure~\ref{fig:si_zeromodes}).
Because $c$ is small against the stiff LJ modes ($\approx43$~eV/{\AA}$^2$), the integer count is unchanged ($n_0=6$): a counting check alone would pass the defective model.
The continuous residual is what catches it: reference LJ $\mathcal{S}=1.000$ ($d=3.9\times10^{-7}$, $n_0=6$, $n_-=0$); anchor modulation $\mathcal{S}=0.45$ ($d=3.9\times10^{-3}$, $n_0=6$, $n_-=0$, gate \texttt{FAIL}).

\clearpage
\section*{Supplementary Note 6: Many-body trimer probe}

\textbf{Identity and target.} 
On the symmetric trimer $A$--$B$--$C$ two exact statements are combined.
(i) The many-body-expansion three-body term $\Delta E_3=E(ABC)-\sum_{\mathrm{pairs}}E+\sum_{\mathrm{atoms}}E$ must decay smoothly and monotonically to $0$ as the apex recedes.
(ii) The force Jacobian must be symmetric: $\langle\mathbf{w},(\mathsf{J}-\mathsf{J}^{\!\top})\mathbf{u}\rangle=0$ for all $\mathbf{u},\mathbf{w}$ (the transverse term).
Both targets are $0$.

\textbf{What the implementation computes.} 
Three-body term: monomer energies and the fixed $AC$ pair are evaluated once; the apex $B$ recedes along $+y$ over a grid of separations ($12$ points to $8$~{\AA} in this demonstration), re-evaluating the trimer and the two moving pairs to form $\Delta E_3(d)$.
The deviation combines the terminal residual $|\Delta E_3^{(N)}|/\max_k|\Delta E_3^{(k)}|$ with a total-variation monotonicity penalty ($\lambda_{\Delta E_3}=0.2$); below a floating-point noise floor an additive model is recognised as trivially vanishing.
Transverse term: for $n_\perp=8$ seeded pairs of random unit vectors the contraction $|\langle\mathbf{w},\mathsf{J}\mathbf{u}\rangle-\langle\mathbf{u},\mathsf{J}\mathbf{w}\rangle|$ is formed matrix-free (two calls per Jacobian--vector product) and averaged ($\lambda_\perp=0.05$); this is the cheap realisation of the Jacobian-symmetry check previously considered too costly to run directly~\cite{pet_oam}.
The two are combined as in the main text (geometric mean of the two sub-scores).

\begin{figure}[htb]
    \centering
    \includegraphics{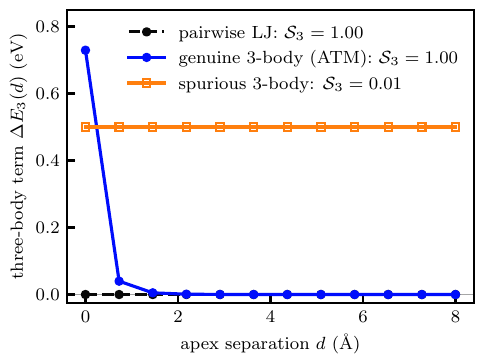}
    \caption{\textbf{Many-body trimer probe.} 
    Three-body term $\Delta E_3(d)$ as the apex $B$ recedes from the $AC$ edge. 
    The pairwise reference (black) is zero to noise; the genuine ATM term (blue) is large at contact but decays smoothly to zero and passes; the spurious term depending only on $|r_{AB}-r_{BC}|$ (orange) plateaus at a finite value and is rejected, although it is the more accurate of the two modulations by both energy and force MAE.}
    \label{fig:si_trimer}
\end{figure}

\textbf{Toy demonstration.} 
Three candidates are compared: (a) the pairwise reference (no three-body term); (b) a genuine three-body term of Axilrod--Teller--Muto (ATM) form, which decays as the apex recedes; and (c) a spurious three-body term $\propto\exp[-((r_{AB}-r_{BC})/0.5)^2]$ that depends only on the difference of the two arms and therefore persists as the apex recedes to infinity.
Against the pairwise reference the genuine ATM modulation has energy MAE $0.24$ and force MAE $1.44$, while the spurious modulation is markedly more accurate (energy MAE $0.16$, force MAE $0.089$).
The probe reports (Supplementary Figure~\ref{fig:si_trimer}): pairwise $\mathcal{S}=1.00$; genuine ATM $\mathcal{S}=1.00$ (three-body sub-score $1.00$); spurious $\mathcal{S}=0.08$ (three-body sub-score $0.01$, rejected despite its small MAE).
All three are conservative, so the transverse sub-score stays $0.99$ throughout.

\clearpage
\section*{Supplementary Note 7: Collinear conservativeness}

\textbf{Identity and target.} 
Along a bond stretch parameterised by the displacement $s$ of one bonded atom along the bond axis $\hat{\mathbf{u}}$, a conservative model satisfies $\mathbf{F}_j\!\cdot\!\hat{\mathbf{u}}=-\mathrm{d}E/\mathrm{d}s$; the target is $0$.

\textbf{What the implementation computes.} 
Following the conservation-deviation diagnostic of MLIP-Arena~\cite{darkside,mliparena}, the routine selects the closest pair as the representative bond, displaces one atom along the bond axis over a symmetric scan ($\pm0.15$~{\AA}; $41$ points in this demonstration), and records the projected force and the energy at each point.
It infers $\mathrm{d}E/\mathrm{d}s$ from a fourth-order central-difference stencil, so truncation error is negligible, and takes the deviation as the mean absolute mismatch $\langle|\mathbf{F}_j\!\cdot\!\hat{\mathbf{u}}+\mathrm{d}E/\mathrm{d}s|\rangle$ over the scan interior, scored with $\lambda=0.05$.

\begin{figure}[htb]
    \centering
    \includegraphics{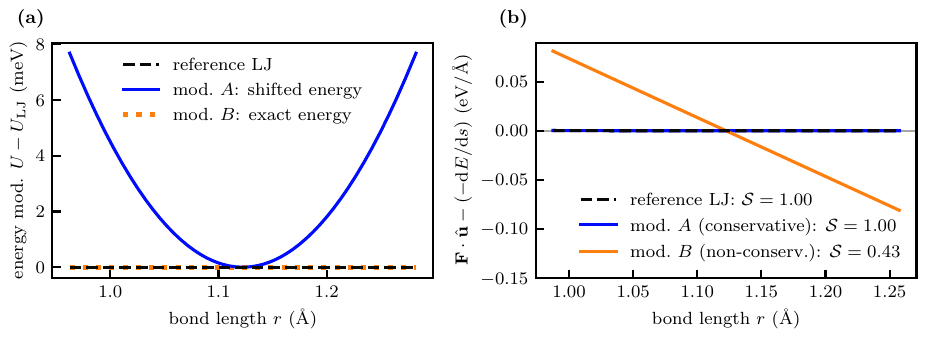}
    \caption{\textbf{Collinear conservativeness.} 
    (a)~The two modulations of the dimer energy: candidate $A$ (blue) carries the parabolic energy offset $\delta(r)$, candidate $B$ (orange, dotted) has exactly the reference LJ energy. 
    Both share the same force (the exact gradient of $A$'s energy) and hence the same force MAE. 
    (b)~Residual $\mathbf{F}\!\cdot\!\hat{\mathbf{u}}-(-\mathrm{d}E/\mathrm{d}s)$ along the scan: zero for the reference and for $A$; equal to $-\delta'(s)$ for $B$, whose forces are not the gradient of its own energy.}
    \label{fig:si_conservativeness}
\end{figure}

\textbf{Toy demonstration.} 
On the analytic dimer two modulations are built with identical forces, hence identical force MAE ($0.015$).
Candidate $A$ is conservative: its energy is shifted by the smooth offset $\delta(r)=c\,(r-r_{\min})^2$ ($c=0.3$) and its forces are the exact gradient of that shifted energy.
Candidate $B$ keeps the exact reference energy (energy MAE $=0$) but is assigned the same forces as $A$; its forces are therefore inconsistent with its own energy by construction, the signature of a direct-force architecture.
The probe reports (Supplementary Figure~\ref{fig:si_conservativeness}): reference LJ $\mathcal{S}=1.00$ ($d=5.9\times10^{-5}$); conservative modulation $A$ $\mathcal{S}=1.00$ ($d=5.9\times10^{-5}$); non-conservative modulation $B$ $\mathcal{S}=0.43$ ($d=4.2\times10^{-2}$).
The two candidates are indistinguishable by force error, and $B$ is strictly the more accurate by energy error, yet only $B$ violates the identity and only $B$ is penalised, with a residual that is precisely the analytic mismatch $-\delta'(s)$.

\clearpage
\section*{Supplementary Note 8: Maxwell--Betti reciprocity}

\textbf{Identity and target.} 
The reciprocal theorem: the force induced at DOF $j$ by a unit displacement of DOF $i$ equals the reciprocal response, $\mathbf{H}_{ij}=\mathbf{H}_{ji}$; the cross-response asymmetry has target $0$.

\textbf{What the implementation computes.} 
For $N_p$ random pairs of Cartesian DOF ($30$ in this demonstration) the routine pushes along $i$ by $\pm\delta$ and reads the induced force at $j$, giving $\mathbf{H}_{ji}$ by central difference, then exchanges the roles for $\mathbf{H}_{ij}$ (four calls per pair).
The deviation is the mean asymmetry $\langle|\mathbf{H}_{ji}-\mathbf{H}_{ij}|\rangle$, scored with $\lambda=0.02$.
It is the canonical-basis, element-wise counterpart of the transverse trimer term: both certify symmetry of the force Jacobian.

\begin{figure}[htb]
    \centering
    \includegraphics{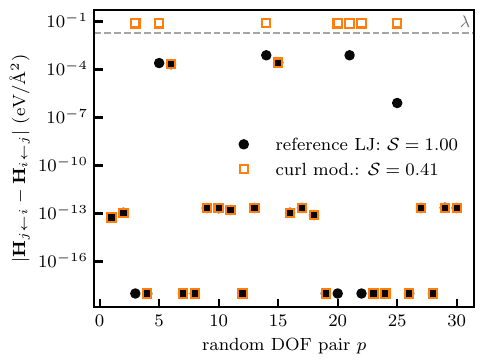}
    \caption{\textbf{Maxwell--Betti reciprocity.} 
    Cross-response asymmetry $|\mathbf{H}_{j\leftarrow i}-\mathbf{H}_{i\leftarrow j}|$ for $30$ random DOF pairs (log scale; dashed line: the probe scale $\lambda=0.02$). 
    The conservative reference (black) never exceeds the finite-difference noise. 
    For the curl modulation (orange) every pair that couples $x$ to $y$ (the plane of the added solenoidal force) jumps above $\lambda$, while its remaining pairs stay at noise, localising the non-conservative coupling.}
    \label{fig:si_betti}
\end{figure}

\textbf{Toy demonstration.} 
The deceptive candidate adds the solenoidal force $\bm{\omega}\times(\mathbf{R}-\mathbf{R}_{\mathrm{cm}})$ with $\bm{\omega}=0.5\,\hat{\mathbf{z}}$: the energy is exactly intact, but the force field is not the gradient of any scalar, so its Jacobian acquires a constant antisymmetric part.
The probe reports (Supplementary Figure~\ref{fig:si_betti}): reference LJ $\mathcal{S}=1.00$ ($d=7.6\times10^{-5}$, largest asymmetry $7.7\times10^{-4}$ corresponding to finite-difference noise); curl modulation $\mathcal{S}=0.41$ ($d=1.8\times10^{-2}$, largest asymmetry $7.8\times10^{-2}$) at a force MAE of only $0.16$.
The per-pair read-out carries structure: the curl about $\hat{\mathbf{z}}$ contributes an antisymmetric part only to $x$--$y$ cross-responses, so exactly those random pairs stand two to three orders of magnitude above the reference noise, while pairs not coupling $x$ to $y$ remain at the noise floor.

\clearpage
\section*{Supplementary Note 9: Strain--position reciprocity}

\textbf{Identity and target.} 
On the extended configuration space of positions and cell strain a conservative model is one scalar $E(\mathbf{R},\bm{\varepsilon})$, and Schwarz's theorem fixes the mixed block: $V\,\partial\sigma_{mn}/\partial R_{j\alpha}=-\,\partial F_{j\alpha}/\partial\varepsilon_{mn}$ (the internal-strain tensor $\Lambda$ of lattice dynamics).
Together with the Voigt symmetry of the strain--strain block $B_{mn}=\partial\sigma_m/\partial\varepsilon_n$, the target is $0$.

\textbf{What the implementation computes.} 
The routine first relaxes positions and cell to the model's own equilibrium.
For $N_p$ random pairs of a Cartesian DOF $j$ and a Voigt component $m$ ($20$ in this demonstration) it central-differences the stress under a displacement of $j$ (left side) and the force under the strain $\varepsilon_m$ (right side), forming the relative per-pair residual defined in the main text, with floor $f=10^{-3}$; $d_\Lambda$ is their mean.
The $B$ block is assembled from $12$ additional stress calls and $d_B=\lVert B-B^{\!\top}\rVert_F/\lVert B\rVert_F$; the score is $\exp[-\tfrac12(d_\Lambda+d_B)/\lambda]$ with $\lambda=0.05$.

\begin{figure}[htb]
    \centering
    \includegraphics{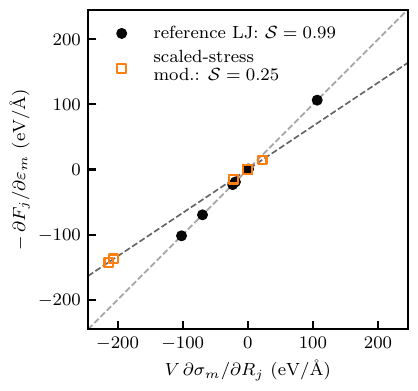}
    \caption{\textbf{Strain--position reciprocity.} 
    The two sides of the mixed-derivative identity for random (DOF, Voigt) pairs: stress response to a displacement (abscissa) against force response to a strain (ordinate). 
    For the reference LJ (black) every pair falls on the identity line; for the scaled-stress modulation (orange) the responsive pairs fall on a line of slope $1/1.5$, the fingerprint of the inflated stress head.}
    \label{fig:si_crossmaxwell}
\end{figure}

\textbf{Toy demonstration.} 
The deceptive candidate returns the exact LJ energy and forces but multiplies its reported stress by $1.5$ (mimicking a mis-scaled stress head), the failure mode of models whose virial is not obtained by differentiating the energy.
Its energy and force MAE are exactly zero.
It violates reciprocity between the two first-derivative heads: the stress response to a displacement is inflated by $1.5$ while the force response to a strain is exact, so every responsive pair carries the relative residual $|1.5x-x|/1.5|x|=1/3$.
The probe reports (Supplementary Figure~\ref{fig:si_crossmaxwell}): reference LJ $\mathcal{S}=0.99$ ($d_\Lambda=4.4\times10^{-4}$, $d_B=2.9\times10^{-4}$); scaled-stress modulation $\mathcal{S}=0.25$ ($d_\Lambda=0.137$, $d_B=8.2\times10^{-5}$).
Note the diagnostic structure: $d_B$ remains at noise (uniform scaling preserves the Voigt symmetry of $B$) so the probe localises the defect specifically to the cross block connecting the force and stress sectors, the coupling that corrupts variable-cell relaxations and NPT dynamics.

\clearpage
\section*{Supplementary Note 10: Stress--gradient consistency}

\textbf{Identity and target.} 
For a periodic cell the stress is the volume-normalised strain derivative of the energy, $\sigma_{ij}=V^{-1}\partial E/\partial\varepsilon_{ij}$; the relative residual between the model's stress head and the finite-difference stress has target $0$.

\textbf{What the implementation computes.} 
On the periodic fcc substrate the routine reads the model's stress once, then for each of the six Voigt strains applies the deformation gradient $\mathsf{I}+\bm{\varepsilon}_k$ with atoms scaled affinely and forms the central-difference component $\sigma_k^{\mathrm{FD}}=[E(+\varepsilon_k)-E(-\varepsilon_k)]/2\delta V$.
The deviation is the relative Frobenius residual $\lVert\bm{\sigma}^{\mathrm{pred}}-\bm{\sigma}^{\mathrm{FD}}\rVert_F/\lVert\bm{\sigma}^{\mathrm{FD}}\rVert_F$, scored with $\lambda=0.05$ ($13$ calls).

\begin{figure}[htb]
    \centering
    \includegraphics{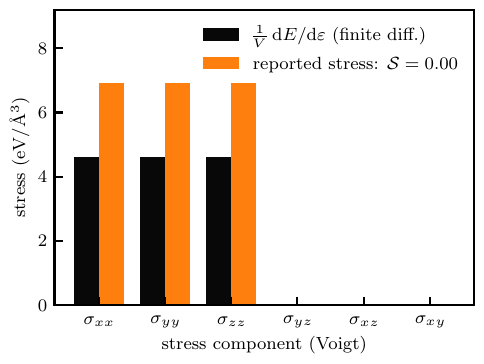}
    \caption{\textbf{Stress--gradient consistency.} 
    Voigt stress components from the finite-difference energy gradient (black) and from the reported stress head (orange) for the scaled-stress modulation. 
    Energy and force MAE are exactly zero, yet the mis-scaled head yields a relative residual of $0.50$ and a score of $0.00$. 
    The shear components vanish identically on the cubic cell, for the head and the gradient alike.}
    \label{fig:si_stress}
\end{figure}

\textbf{Toy demonstration.} 
The same scaled-stress candidate as in Supplementary Note~9 (stress head $\times1.5$) is used in this probe; its energy and force MAE are exactly zero.
The probe reports (Supplementary Figure~\ref{fig:si_stress}): reference LJ $\mathcal{S}=1.000$ ($d=2.0\times10^{-5}$); scaled-stress modulation $\mathcal{S}=0.00$ ($d=0.50$, gate \texttt{FAIL}).

\clearpage
\section*{Supplementary Note 11: Time-reversibility}

\textbf{Identity and target.} 
An NVE trajectory integrated forward $N$ velocity-Verlet steps, then backward $N$ steps after reversing the momenta, returns to its origin; the per-atom RMS return displacement $d_{\mathrm{rt}}$ has target $0$.

\textbf{What the implementation computes.} 
The routine sets seeded Maxwell--Boltzmann velocities, propagates $N=100$ steps at 1~fs, reverses all momenta, propagates back, and reports $d_{\mathrm{rt}}$, scored with $\lambda=0.001$~{\AA}; it separately records the secular energy drift as the slope of the total energy along the trajectory. 
For this demonstration the temperature is $900$~K.
In exact arithmetic velocity Verlet is time-reversible for any force that is a deterministic function of the current positions, whether smooth, kinked, discontinuous or non-conservative. 
The round trip is therefore blind to non-conservativeness by construction, and it isolates what breaks that assumption: a force that depends on anything other than the current positions.
This covers hidden internal state, such as a neighbour list whose rebuild history differs between the passes, and non-deterministic force evaluation, for example unordered parallel reductions that return slightly different forces for identical positions. 
In finite precision the return is further limited by round-off, which the dynamics amplifies more strongly on a stiff or rough surface. 
A force discontinuity contributes only when round-off places the forward and backward passes on opposite sides of it, so the probe is not a reliable detector of discontinuities; bond smoothness is the axis for those.

\begin{figure}[htb]
    \centering
    \includegraphics{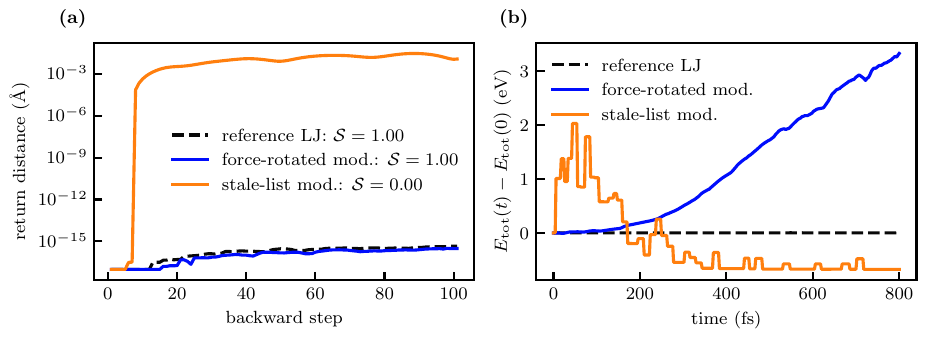}
    \caption{\textbf{Time-reversibility.} 
    (a)~Distance from the stored forward trajectory during backward integration (log scale). 
    The reference LJ (black) and the non-conservative force-rotated modulation (blue) retrace to machine precision (in exact arithmetic, velocity Verlet is reversible for any deterministic, position-dependent force). 
    The stale-neighbour-list modulation (orange) departs by thirteen orders of magnitude the moment its rebuild history differs between the passes. 
    (b)~The complementary drift read-out over a longer NVE run.}
    \label{fig:si_reversibility}
\end{figure}

\textbf{Toy demonstration.} 
The deceptive candidate is a stale-neighbour-list LJ, the classic implementation shortcut: plain truncated LJ ($r_c^{\mathrm{pair}}=1.75$, skin 0.1) whose pair list is rebuilt only once an atom has moved by half the skin since the last rebuild, so the force at a configuration depends on the trajectory history. 
The control candidate rotates every force vector by a fixed $0.05$~rad; it is therefore non-conservative (force MAE $0.42$) but deterministic and position-dependent, with an exact energy. 
The probe reports (Supplementary Figure~\ref{fig:si_reversibility}): reference LJ $d_{\mathrm{rt}}=4.6\times10^{-16}$~{\AA}, $\mathcal{S}=1.000$; force-rotated modulation $d_{\mathrm{rt}}=3.1\times10^{-16}$~{\AA}, $\mathcal{S}=1.000$; stale-list modulation $d_{\mathrm{rt}}=1.1\times10^{-2}$~{\AA}, thirteen orders of magnitude above the reference, $\mathcal{S}=0.00$ on the default scale. 
The divergence begins abruptly at the first backward step at which the list-rebuild history differs from the forward pass, visible as the step in panel~(a).
The drift read-out completes the picture: over an $800$~fs NVE run the force-rotated model, invisible to the round trip, gains $+3.3$~eV with an energy MAE of exactly zero, while the stale-list model both drifts ($-0.67$~eV) and is irreversible. 
The two read-outs therefore separate non-conservativeness (drift without irreversibility) from hidden state (irreversibility).

\clearpage
\section*{Supplementary Note 12: Configurational temperature}

\textbf{Identity and target.} 
The configurational temperature $k_BT_{\mathrm{conf}}=\langle\lVert\mathbf{F}\rVert^2\rangle/\langle\operatorname{Tr}\mathbf{H}\rangle$~\cite{hutchinson1989,rugh1997,butler1998} equals the kinetic temperature at Boltzmann equilibrium: $T_{\mathrm{conf}}/T_{\mathrm{kin}}\to1$.

\textbf{What the implementation computes.} 
The routine relaxes to the model's own minimum, runs a Langevin-thermostatted trajectory ($30$~K, $400$ warm-up and $1{,}100$ production steps at $1.0$~fs, sampling every 10th step), and averages both channels over snapshots: the trace is a matrix-free Hutchinson estimate~\cite{hutchinson1989} over four seeded $\pm1$ vectors per snapshot, and the kinetic temperature uses the full $3N$ DOF.
The deviation is $d_T=|T_{\mathrm{conf}}/T_{\mathrm{kin}}-1|$ with $\lambda=1$.
Being a finite-time average, the ratio carries irreducible sampling noise; the exact target holds in the long-trajectory limit.

\begin{figure}[htb]
    \centering
    \includegraphics{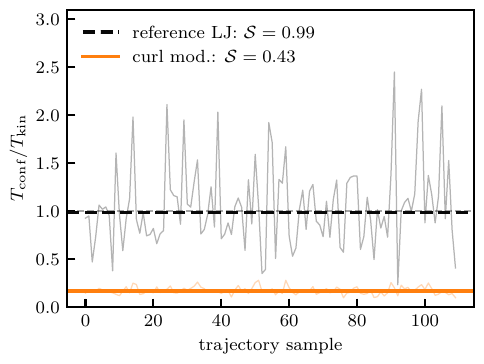}
    \caption{\textbf{Configurational temperature.} 
    Per-snapshot ratio $T_{\mathrm{conf}}/T_{\mathrm{kin}}$ (thin) and the scored ratio of time averages (bold) along the thermostatted trajectory. 
    The reference LJ (black) centres near the ideal value $1$ within finite-sampling noise; the curl modulation (orange), which samples no Boltzmann distribution, sits systematically at $0.17$.}
    \label{fig:si_configtemp}
\end{figure}

\textbf{Toy demonstration.} 
The deceptive candidate adds the solenoidal force $\bm{\omega}\times(\mathbf{R}-\mathbf{R}_{\mathrm{cm}})$, $\bm{\omega}=1.0\,\hat{\mathbf{z}}$: the energy is untouched, but the dynamics samples no Boltzmann distribution, so its two temperatures cannot agree.
The probe reports (Supplementary Figure~\ref{fig:si_configtemp}): reference LJ $\mathcal{S}=0.99$ (ratio $0.99$, off unity by the finite-sampling noise discussed above); curl modulation $\mathcal{S}=0.43$ (ratio $0.17$) at a force MAE of $0.34$.
The reference's per-snapshot ratio fluctuates strongly but centres near unity; the non-conservative model sits far below it across the entire trajectory, a systematic, not a statistical, deviation.

\clearpage
\section*{Supplementary Note 13: Virial theorem}

\textbf{Identity and target.} 
The Clausius virial theorem holds for any bound classical system in a stationary state: $\rho_{\mathrm{vir}}=-\langle\sum_i(\mathbf{R}_i-\mathbf{R}_i^{\mathrm{eq}})\cdot\mathbf{F}_i\rangle/[(3N-6)\,k_BT^{\mathrm{int}}_{\mathrm{kin}}]\to1$.

\textbf{What the implementation computes.} 
On a Langevin trajectory ($30$~K, $400$ warm-up and $1{,}100$ sampling steps at $1.0$~fs) the routine accumulates, at every fourth snapshot, the virial numerator from the forces and the displacements $\mathbf{R}_i-\mathbf{R}_i^{\mathrm{eq}}$ from the model's own relaxed minimum $\mathbf{R}^{\mathrm{eq}}$, from which the trajectory starts. 
In a stationary state, where every $\langle\mathbf{F}_i\rangle$ vanishes, this has the same expectation as the centre-of-mass form but drops the large zero-mean term $\sum_i(\mathbf{R}_i^{\mathrm{eq}}-\mathbf{R}_{\mathrm{cm}})\cdot\mathbf{F}_i$. 
The internal kinetic energy is obtained after projecting the rigid translation and rotation out of the velocities, since the six unconfined DOF carry kinetic energy but no confining virial. 
The deviation is $\lvert\rho_{\mathrm{vir}}-1\rvert$ with $\lambda=1$.

\begin{figure}[htb]
    \centering
    \includegraphics{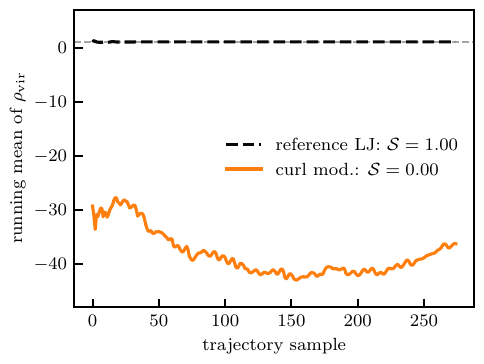}
    \caption{\textbf{Virial theorem.} 
    Running mean of the virial ratio $\rho_{\mathrm{vir}}$ along the thermostatted trajectory. 
    The reference LJ (black) settles onto the ideal value $1$; for the curl modulation (orange) the solenoidal force pumps energy and angular momentum into the cluster, the state never becomes stationary, and the running mean diverges to $-34$ (scored ratio $\rho_{\mathrm{vir}}=-33.9$).}
    \label{fig:si_virial}
\end{figure}

\textbf{Toy demonstration.} 
The same curl modulation ($\bm{\omega}=0.4\,\hat{\mathbf{z}}$) performs net work around closed loops: it continuously pumps energy and angular momentum into the cluster, so the sampled state is not stationary and the force--position correlation reverses sign.
The probe reports (Supplementary Figure~\ref{fig:si_virial}): reference LJ $\mathcal{S}=1.00$ ($\rho_{\mathrm{vir}}=1.0005$); curl modulation $\mathcal{S}=0.00$ ($\rho_{\mathrm{vir}}=-33.9054$) at a force MAE of $0.13$.
The reference's running mean settles onto unity within a few dozen samples; the non-conservative model diverges without bound.

\clearpage
\section*{Supplementary Note 14: Per-mode equipartition}

\textbf{Identity and target.} 
Near a minimum every stable normal mode satisfies $R_\alpha=\omega_\alpha^2\langle Q_\alpha^2\rangle/k_BT\to1$, tying the static stiffness of each mode to the amplitude with which the dynamics populates it.

\textbf{What the implementation computes.} 
The curvature channel assembles the finite-difference Hessian ($6N$ calls), mass-weights and diagonalises it, keeping the stable modes (discarding the six zero modes and any imaginary ones).
The dynamics channel runs a low-temperature Langevin trajectory ($30$~K, $400$ warm-up and $1{,}100$ sampling steps at $1.0$~fs) and, after an equilibration window, accumulates $\langle Q_\alpha^2\rangle$ from mass-weighted displacements projected onto each mode.
The full spectrum $\{R_\alpha\}$ is the mode-resolved diagnostic; since each per-mode ratio carries irreducible sampling scatter, the scalar deviation is the sampling-robust $d_{\mathrm{eq}}=|\operatorname{median}_\alpha R_\alpha-1|$, scored with $\lambda=1.0$.

\begin{figure}[htb]
    \centering
    \includegraphics{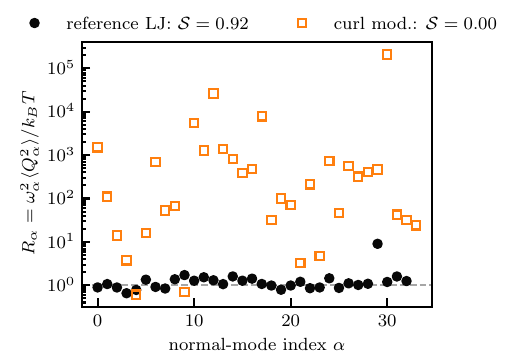}
    \caption{\textbf{Per-mode equipartition.} 
    Ratio $R_\alpha=\omega_\alpha^2\langle Q_\alpha^2\rangle/k_BT$ for every stable normal mode (log scale). 
    The reference LJ (black) clusters on the ideal value $1$; the curl modulation (orange) scatters mode by mode across more than five decades ($R_{\mathrm{median}}=161.09$) at a force MAE of only $0.084$.}
    \label{fig:si_equipartition}
\end{figure}

\textbf{Toy demonstration.} 
The curl modulation (here $\bm{\omega}=0.25\,\hat{\mathbf{z}}$, force MAE $0.084$) leaves the curvature exact but drives the dynamics off the Boltzmann distribution, so stiffness and population no longer agree mode by mode.
The probe reports (Supplementary Figure~\ref{fig:si_equipartition}): reference LJ $\mathcal{S}=0.92$ ($R_{\mathrm{median}}=1.08$ over $33$ modes), the spectrum collapses onto unity within sampling noise; curl modulation $\mathcal{S}=0.00$ ($R_{\mathrm{median}}=161.09$), with individual modes overpopulated by up to five orders of magnitude.

\clearpage
\section*{Supplementary Note 15: Bond-deformation smoothness}

\textbf{Identity and target.} 
Along a bond scan from compression through dissociation a physical surface is a smooth single well: no localised force jumps or kinks, a single zero crossing of the projected force, a single energy minimum, low force tortuosity~\cite{bsct}.
A perfectly smooth well gives every penalty zero and score $1$.

\textbf{What the implementation computes.} 
The routine scans the closest pair from moderate compression (capped at a fraction of the bond length, avoiding the repulsive wall) into the dissociation tail ($161$ points in this demonstration), recording projected force and energy.
It forms the four penalties defined in the main text: (i) the force-spike ratio (entering only above the baseline $\rho_0=4$, at weight $w=\tfrac14$), (ii) excess force sign reversals, (iii) excess energy minima, and (iv) force tortuosity and scores their weighted sum with $\lambda=1$.

\begin{figure}[htb]
    \centering
    \includegraphics{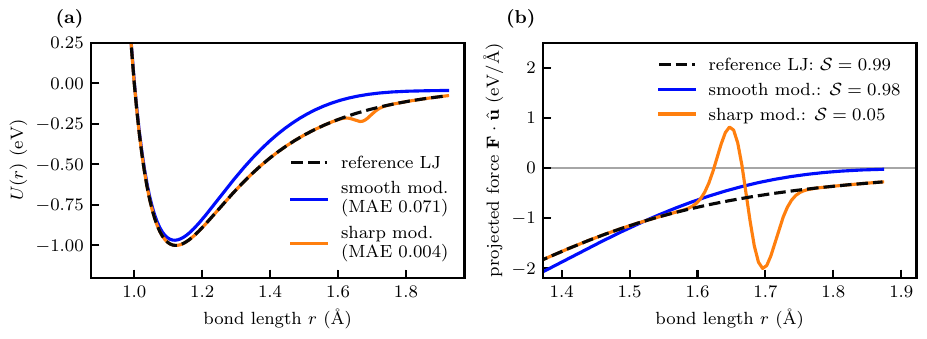}
    \caption{\textbf{Bond-deformation smoothness.} 
    (a)~Dimer potential wells: reference LJ (black, dashed), smooth broad modulation (blue, energy MAE $0.071$), sharp localised modulation (orange, energy MAE $0.0039$). 
    (b)~Projected force along the bond in the dissociation tail: the smooth modulation tracks a monotone force and scores $0.98$; the sharp modulation introduces two extra force reversals and a spurious minimum and scores $0.05$.}
    \label{fig:si_smoothness}
\end{figure}

\textbf{Toy demonstration.} 
On the analytic dimer the reference is compared against two Gaussian modulations of the well.
The smooth modulation adds a broad, low bump, $(+0.12)\!\cdot\!\exp[-((r-r_{\min}-0.40)/0.35)^2]$; the sharp modulation adds a narrow dip in the dissociation tail, $(-0.06)\!\cdot\!\exp[-((r-r_{\min}-0.55)/0.035)^2]$.
The calibration is deliberate: the sharp modulation has an energy MAE of $0.0039$, eighteen times smaller than the smooth modulation's $0.071$.
By a fixed-test-set metric it is by far the better model.
The probe reports (Supplementary Figure~\ref{fig:si_smoothness}): reference LJ $\mathcal{S}=0.99$; smooth modulation $\mathcal{S}=0.98$ ($1$~force crossing, $0$ spurious minima, $d=0.02$); sharp modulation $\mathcal{S}=0.05$ ($3$ force crossings, $1$ spurious minimum, $d=3.05$).
The localised dip creates two spurious force reversals and a false minimum that a trajectory would encounter as a trap in the dissociation tail, nearly invisible to the averaged energy error.

\clearpage
\section*{Supplementary Note 16: Summary of the validation}

Supplementary Table~\ref{tab:si_summary} collects the scores of all fourteen probes on the reference LJ potential and on the modulation designed to defeat each.
Two uniform observations emerge.
First, the reference scores near $1$ everywhere: no probe penalises a correct potential, and the small departures of the trajectory-based probes ($0.92$--$1.00$) quantify their finite-sampling floor, not a defect.
Second, every adversarial modulation is detected, including three whose energy and force errors are exactly zero and four that are strictly more accurate than the alternative modulation the probe accepts.

\begin{table}[htb]
    \centering
    \caption{\textbf{Validation summary.} 
    Score ($1=$ ideal) on the reference LJ potential and on the adversarial modulation for each probe, with the adversary's accuracy against the reference (energy MAE per atom in eV, force MAE in eV/{\AA}).}
    \label{tab:si_summary}
    \begin{tabular}{l l l c c l}
        \toprule
        S.-Note & Probe & Adversarial modulation & $\mathcal{S}$(LJ) & $\mathcal{S}$(mod) & Accuracy of the adversary \\
        \midrule
        2  & Equivariance          & absolute-orientation term      & $1.00$ & $0.00$ & $E$-MAE $0.024$ (energy rotation-invariant) \\
        3  & Parity stress rule    & parity-odd $\sigma_{xy}$ leak  & $1.00$ & $0.00$ & $E$- and $F$-MAE exactly $0$ \\
        4  & Representation        & egg-box grid term              & $1.00$ & $0.00$ & $E$-MAE $0.015$, $F$-MAE $0.032$ \\
        5  & Hessian null space    & absolute-position spring       & $1.00$ & $0.45$ & $E$-MAE $0.012$, $F$-MAE $0.16$ \\
        6  & Trimer (many-body)    & non-vanishing 3-body term      & $1.00$ & $0.08$ & more accurate than accepted ATM mod. \\
        7  & Conservativeness      & forces of a different energy   & $1.00$ & $0.43$ & $F$-MAE $0.015$ (= accepted mod.\ $A$), $E$-MAE $0$ \\
        8  & Maxwell--Betti        & solenoidal (curl) force        & $1.00$ & $0.41$ & $F$-MAE $0.16$, $E$-MAE $0$ \\
        9  & Strain--position      & stress head $\times1.5$        & $0.99$ & $0.25$ & $E$- and $F$-MAE exactly $0$ \\
        10 & Stress--gradient      & stress head $\times1.5$        & $1.00$ & $0.00$ & $E$- and $F$-MAE exactly $0$ \\
        11 & Reversibility         & stale neighbour list           & $1.00$ & $0.00$ & $E$-MAE $0.12$; $\Delta_{\mathrm{rt}}$ up $13$ orders \\
        12 & Config.\ temperature  & solenoidal (curl) force      & $0.99$ & $0.43$ & $F$-MAE $0.34$, $E$-MAE $0$ \\
        13 & Virial theorem        & solenoidal (curl) force        & $1.00$ & $0.00$ & $F$-MAE $0.13$, $E$-MAE $0$ \\
        14 & Equipartition         & solenoidal (curl) force        & $0.92$ & $0.00$ & $F$-MAE $0.084$, $E$-MAE $0$ \\
        15 & Smoothness            & sharp localised modulation     & $0.99$ & $0.05$ & $E$-MAE $0.0039$ ($18\times$ better than accepted mod.) \\
        \bottomrule
    \end{tabular}
\end{table}

\clearpage
\section*{Supplementary Note 17: Per-probe results of the fine-tuning study}

This note records the full per-probe output behind Figure~3 of the main text, together with the accuracy measurements and the molecular-dynamics test that the certification is compared against.

{\begin{center} \textbf{Models and data} \end{center}}

The base model is the universal \textsc{MACE-MP-0} (small) potential (UMLIP).
It is fine-tuned to the sulfur-vacancy jump in layered MoS$_2$ by three protocols: naive fine-tuning, in which all weights are free; LoRA fine-tuning, in which low-rank adapter matrices are trained while the original weights stay frozen; and multi-head replay fine-tuning, in which one head is trained on a subset of the universal training data and a second on the fine-tuning set, simultaneously.
The fine-tuning dataset comprises $572$~first-principles data points distributed roughly uniformly over $30$~bins along the configuration coordinate of the jump, with a minimum of $14$ points per bin; computational details of the underlying DFT calculations are given in the methods section of our previous publication~\cite{atk}.
All four models are certified on molybdenum substrates, so that the substrates are drawn from the fine-tuning chemistry but from none of its configurations.

{\begin{center} \textbf{Per-probe certification} \end{center}}

\begin{figure*}[htb]
    \centering
    \includegraphics[width=\textwidth]{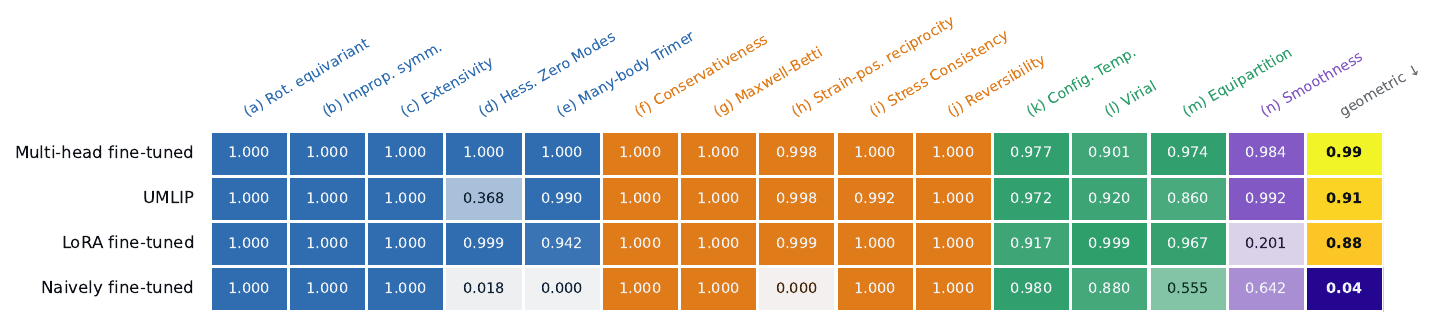}
    \caption{\textbf{Score matrix across fine-tuned models and probes.}
    Rows are the fine-tuned \textsc{MACE-MP-0} (small) models and the universal model itself, columns the fourteen probes grouped by family; colour encodes the normalised score $\mathcal{S}\in[0,1]$ ($1=$ ideal).
    The probes were evaluated on molybdenum substrates.}
    \label{fig:si_heatmap}
\end{figure*}

Supplementary Figure~\ref{fig:si_heatmap} lists every probe score.
The three protocols are not distinguished by the probes that the architecture guarantees: all four models return machine-precision scores on equivariance, the improper-symmetry stress rule, representation invariance, collinear conservativeness, stress--gradient consistency, Maxwell--Betti reciprocity and time-reversibility.
Fine-tuning cannot break what \textsc{MACE} builds in, and the certification correctly reports that it has not.

The protocols separate sharply on the probes that the architecture does not guarantee, and the naive model fails five of them at once.
Its many-body trimer score collapses to $0.000$: the three-body term no longer decays as the apex recedes, so the model has acquired a spurious interaction that survives at arbitrary separation.
Its strain--position reciprocity also falls to $0.000$, but here the score is dominated by the relaxation that precedes it: for the naive model the joint relaxation of positions and cell does not converge within its $200$ steps and runs away to stresses of order $10^{9}$~eV/{\AA}$^3$, so the finite differences are taken on an unphysical cell.
Since \textsc{MACE} obtains the stress by differentiating its energy, this score records the runaway under strain rather than a disagreement between separate force and stress heads.
The Hessian null space ($0.018$), the equipartition probe ($0.555$) and the bond-deformation smoothness ($0.642$) follow.
The pattern is coherent: unconstrained fine-tuning on a narrow configuration set preserved every hard-wired symmetry and destroyed the transferable structure of the surface that the narrow set does not constrain~\cite{grunert2025,hanseroth2025optimizing}.
LoRA and multi-head fine-tuning avoid this collapse, and both improve on the base model's weakest column, the Hessian null space, where the UMLIP itself scores only $0.368$.
The certificate is not uniformly favourable to the parameter-efficient protocols either: LoRA records the lowest smoothness score of the four ($0.201$), a genuine roughness in the bond scan that is nevertheless mild enough not to destabilise dynamics (see below).

\newpage
{\begin{center} \textbf{Accuracy on the fine-tuning target} \end{center}}

Supplementary Table~\ref{tab:si_ft_mae} gives the conventional accuracy measurement: mean absolute errors on a held-out set of $200$ MoS$_2$ configurations drawn from the same MD simulation trajectory as the fine-tuning data, but disjoint from it, so that no test configuration was seen during fine-tuning.
By this measurement all three fine-tuned models improve substantially on the universal model, and the naively fine-tuned model attains the lowest force error of the three.
Nothing in this table, nor in the predicted jump barrier of Figure~3a of the main text, identifies the naive model as the defective one.
This is the litmus test of the main text realised on a real workflow rather than on an analytic construction: the two criteria a practitioner would normally apply both rank the naive model first, and the certification ranks it last by a factor of twenty.

The energy column requires one remark.
Fine-tuning re-references the energy scale, which is why all three fine-tuned models drop to the meV range.

\begin{table}[htb]
    \centering
    \caption{\textbf{Accuracy on the fine-tuning target.}
    Mean absolute force and energy errors of the universal model and the three fine-tuned models on a held-out test set of $200$ MoS$_2$ configurations.}
    \label{tab:si_ft_mae}
    \begin{tabular}{l c c}
        \toprule
        Model & MAE force (eV/{\AA}) & MAE energy (eV/atom) \\
        \midrule
        \textsc{MACE-MP-0} (small), universal         & $0.252$ & $807.509$ \\
        Naively fine-tuned                            & $0.055$ & $0.011$ \\
        LoRA fine-tuned                               & $0.060$ & $0.002$ \\
        Multi-head fine-tuned                         & $0.151$ & $0.003$ \\
        \bottomrule
    \end{tabular}
\end{table}

{\begin{center} \textbf{Behaviour outside the fine-tuning domain} \end{center}}

The certification is a statement about the surface, not about the fine-tuning target, so its prediction is tested outside the fine-tuning chemistry.
We use octathiocane (S$_8$), a molecular crystal built entirely of sulfur rings~\cite{zeng2021}: chemically adjacent to the fine-tuning system, which contains sulfur, but present in none of its configurations.
Two measurements are made on it.

First, accuracy: all four models recalculate forces on ten configurations taken from an ab initio molecular-dynamics trajectory of the S$_8$ crystal at $300$~K (Supplementary Table~\ref{tab:si_ft_ood}).
Every fine-tuned model is less accurate here than the universal model it started from, as is expected when a universal potential is specialised, but the ordering among the three fine-tuned models does not single out the naive one: its out-of-domain force error ($0.50$~eV/{\AA}) is the smallest of the three, smaller than that of the multi-head ($0.58$) and the LoRA model ($0.79$).
Out-of-domain accuracy therefore fails to identify the defective model just as in-domain accuracy does.

Second, stability: each model runs $10{,}000$ steps of Nos\'e--Hoover-chain dynamics on the same S$_8$ crystal at $300$~K and $1$~fs, starting from identical seeded Maxwell--Boltzmann velocities with the centre of mass fixed.
The universal, LoRA and multi-head models complete all $10{,}000$ steps and remain thermostatted between $240$ and $380$~K throughout.
The naive model heats to $1{,}960$~K within the first $100$~steps, overflows the thermostat integration and returns non-finite coordinates at step~$140$, i.e.\ after $1.4$~\% of the requested trajectory.
Only the certification, which needs no reference data and consumed $4{,}256$ single-point evaluations in $76$~s of wall time, anticipated this outcome.

\begin{table}[htb]
    \centering
    \caption{\textbf{Behaviour outside the fine-tuning chemistry.}
    Mean absolute force error and mean relative force error of each model against reference forces on ten configurations of S$_8$ crystal drawn from an AIMD trajectory at $300$~K.
    Outcome of $10{,}000$ steps of Nos\'e--Hoover-chain dynamics on the same crystal at $300$~K.}
    \label{tab:si_ft_ood}
    \begin{tabular}{l c c c}
        \toprule
        Model & MAE force (eV/{\AA}) & Rel.\ force error & MD outcome ($10{,}000$ steps at $300$~K) \\
        \midrule
        \textsc{MACE-MP-0} (small), universal & $0.175$ & $0.62$ & completed, $240$--$380$~K \\
        Naively fine-tuned                    & $0.501$ & $1.77$ & non-finite at step $140$, ($1{,}960$~K by step $100$) \\
        LoRA fine-tuned                       & $0.791$ & $2.79$ & completed, $240$--$380$~K \\
        Multi-head fine-tuned                 & $0.578$ & $2.04$ & completed, $240$--$380$~K \\
        \bottomrule
    \end{tabular}
\end{table}

Across the three protocols, the same base model and the same $572$ training points produce certificates from $0.04$ to $0.99$.
In-domain force error, the predicted physical observable and out-of-domain force error all rank the naively fine-tuned model at or near the top of the fine-tuned series; the reference-free certification is the only measurement reported here that ranks it last, and it is the ranking that the molecular-dynamics test confirms.
Because the certification requires no reference calculation and completes in about a minute per model, it can be evaluated at every fine-tuning checkpoint as a physics regression test.

\bibliography{bibliography}